\documentclass[useAMS,usenatbib]{mn2e}
\usepackage{epsfig}

\newcommand{\co}{\mbox{$^{12}$CO}}
\newcommand{\coa}{\mbox{$^{13}$CO}}
\newcommand{\cob}{\mbox{C$^{18}$O}}

\newcommand{\kms}{\mbox{km s$^{-1}$}}
\newcommand{\htwo}{\mbox{H$_2$}}
\newcommand{\nhtwo}{\mbox{N($H_2$)}}
\newcommand{\wco}{\mbox{$W_{CO}$}}
\newcommand{\xco}{\mbox{$X_{CO}$}}
\newcommand{\ncoa}{\mbox{N($^{13}$CO)}}
\newcommand{\cc}{\mbox{cm$^{-3}$}}
\newcommand{\cmsq}{\mbox{cm$^{-2}$}}
\newcommand{\msun}{\mbox{M$_\odot$}}

\newcommand{\vlsr}{\mbox{$V_{LSR}$}}
\newcommand{\Av}{\mbox{$A_v$}}
\newcommand{\etal}{\mbox{et~al.~}}
\begin{document}

\title[CO Abundance Variations in the Orion Molecular Cloud]{CO Abundance Variations in the Orion  Molecular Cloud}
\author[F. Ripple, M.H. Heyer, R. Gutermuth, R.L. Snell, C.M. Brunt]
{F. Ripple$^1$, M. H. Heyer$^1$, R. Gutermuth$^1$, R.L. Snell$^1$, C.M. Brunt$^2$\\
$^{1}$Department of Astronomy, University of Massachusetts, Amherst, MA 01003, USA\\
$^{2}$School of Physics, University of Exeter, Exeter, EX4 4QL, UK
}
\date{\today}

\maketitle

\label{firstpage}

\begin{abstract}
Infrared stellar photometry from the Two Micron All Sky Survey (2MASS) and  spectral line imaging observations of  \co\ and \coa\ J = 1-0 line emission from the Five College Radio Astronomy Observatory (FCRAO) 14m telescope are analysed 
%to construct independent measures of the \htwo\ column density distribution
 to assess the variation of the CO abundance with physical conditions throughout the 
Orion A and Orion B molecular clouds. Three distinct \Av\ regimes are identified in which the ratio between 
 the \coa\ column density and visual extinction changes corresponding to the photon dominated envelope, the strongly self-shielded interior, and the cold, dense volumes of the clouds.  
Within the strongly self-shielded interior of the Orion A cloud, the \coa\ abundance varies by 100\% with a peak 
value located near regions of enhanced star formation activity.  The effect of CO depletion onto the ice mantles 
of dust grains is limited to regions with \Av\ $>$ 10 mag and gas temperatures less than $\sim$20 K 
as predicted by chemical models that consider thermal-evaporation to desorb molecules from grain surfaces.  

Values of the molecular mass of each cloud are independently derived from the distributions of \Av\ and \coa\ column densities with a constant \coa-to-\htwo\ abundance over various extinction ranges.  Within the strongly self-shielded interior of the cloud (\Av $>$ 3 mag), \coa\ provides a reliable tracer of \htwo\ mass with the exception of the cold, dense volumes where depletion is important.  However, owing to its reduced abundance, \coa\ does not trace the \htwo\ mass that resides in the extended cloud envelope, which comprises 40-50\% of the molecular mass of each cloud.  
 The implied CO luminosity to mass ratios, $M/L_{CO}$, 
are 3.2 and 2.9 for Orion A and Orion B respectively, which are comparable to the value (2.9), derived from $\gamma$-ray observations of the Orion region. 
%These comparisons point to a large reservoir of molecular gas mass residing within the low extinction envelope that is not traced by CO observations
Our results emphasize the need to consider local conditions when applying CO observations to derive \htwo\ column densities. 
\end{abstract}

\section{Introduction}                                               
The mass and column density of interstellar molecular clouds impact their evolution,  chemistry, and the production of newborn stars within their domain.   Therefore, accurate measures of these key properties are essential to our understanding of this important interstellar gas phase.   The millimeter and submillimeter rotational transitions of carbon monoxide (\co) and its isotopologues, \coa\ and \cob, are the most widely used  tracers of molecular hydrogen (\htwo) that offer several advantages over other methods such as dust extinction, dust emission, and gamma rays.  These include the ability to construct high angular resolution views 
of distant interstellar molecular clouds throughout the Milky Way and other galaxies, and with sufficient spectral resolution,
the provision of critical kinematic information such as the velocity centroid and velocity dispersion for a given line of sight.   
Measurements of the optically thick \co\ J=1-0 line are routinely made in investigations of star formation processes in galaxies  \citep{young95, omont07, leroy09}.  The application of this high opacity line as a tracer of molecular hydrogen relies on the empirical correlation between \co\ velocity integrated intensity, W$_{CO}$, and $\gamma$-ray 
intensity that is produced from the interaction of cosmic rays with hydrogen protons within the Milky Way \citep{bloemen1986, strong96,hunter97}.  The corresponding 
CO to \htwo\ conversion factor, $X_{CO}$=\nhtwo/W$_{CO}$ is 1.9$\times$10$^{20}$ cm$^{-2}$ (K km s$^{-1}$)$^{-1}$ \citep{strong96}.  A similar procedure has been followed using the dust emission to trace gas column density with
derived values of
 $X_{CO}$ ranging from 1.8-2.5$\times$10$^{20}$ cm$^{-2}$ (K km s$^{-1}$)$^{-1}$ \citep{dame2001,ade2011}.

Emission from  \coa\ offers another measure of the molecular gas distribution. Owing to 
its lower opacity, it is straightforward to derive the \coa\  column density under the assumption of local thermodynamic 
equilibrium (LTE), using the optically thick \co\ line to estimate the excitation temperature \citep{dickman1978}.
However, to derive the more relevant \htwo\ column density and mass, knowledge of the \coa\ 
abundance is required.  Most studies adopt a constant 
fractional abundance of $\sim$2$\times$10$^{-6}$ throughout the cloud based on previous investigations of nearby dark clouds, which
examined the relationship between \coa\ column density and visual extinction \citep{dickman1978, frerking82}.
These earlier studies investigated the \ncoa-\Av\ relationship in cold, dark clouds with limited star formation activity.

 In this paper, we examine the relationship between carbon monoxide and near infrared extinction in the Orion molecular cloud that is producing massive stars and presents a broad range of environmental conditions. In Section 2, we describe the near-infrared and molecular line data used in this study. The procedures used to derive extinctions from the 2MASS colours and the \coa\ column densities from the molecular line imaging are summarized in Section 3.  In Section 4 the results of point-to-point comparison of the \coa\ column density and \co\ integrated intensity to visual extinction relation are presented. The interpretation of these relationships within the context of photon-dominated regions (PDRs) models, the calculation of relative abundances, and the derivation of cloud masses are described in Section 5. 

\section{Data}
\subsection{Millimeter Spectroscopic Imaging}
Observations of \co\ and \coa\ J=1-0 emission were carried out with the FCRAO 14m telescope using the 32 pixel SEQUOIA array \citep{erickson99} between 2005 and 2006. Owing to the broad bandwidth of the HEMT amplifiers, the \co\ and \coa\ lines were observed simultaneously enabling excellent positional registration and calibration.  The backends were comprised of 64 autocorrelation spectrometers
each with 25 MHz bandwidth.
No smoothing was applied to the autocorrelation function so the 
spectral resolution was 29.5 kHz per channel corresponding to 0.077 \kms\ and 0.080 \kms\ at the line frequencies of 
\co\ (115.271202 GHz) and \coa\ (110.201353 GHz) respectively. 
System temperatures ranged from 350-500 K (\co) and 150-300 K (\coa).
The FWHM beam size of the antenna 
at the observed frequencies are 45\arcsec\ (115 GHz) and 47\arcsec\
(110 GHz).    
The main beam efficiencies  at these frequencies
are 0.45 and 0.48 respectively as determined  from measurements of
Jupiter.  

The areas of the Orion A and Orion B clouds were divided into 
submaps.   A given  submap of \co\ and \coa\ emission was constructed 
by scanning the telescope at an angle $\theta_s=14.04$  degrees
relative to the azimuth direction. This choice of scanning angle ensures even coverage of the sky 
along a single scan.   Subsequent scans are orthogonally offset by the entire width of the array to extend
the sky coverage. We also employed scanning angles of $\theta_s+90$.  A basic observational block was
comprised of a sequence of scans either using $\theta_s$ or $\theta_s+90$.  Submaps were observed so that adjacent blocks overlap.  The scanning angle for each
block was set to be opposite to that of its neighboring block.
The angular coverage spanned  22.2 deg$^2$ for Orion A and 17.4 deg$^2$ for Orion B.  The 1$\sigma$ errors
due to thermal noise (main beam temperature units) within 0.2 \kms\ wide channels are 2.0 K (Orion A) and 2.1 K (Orion B) for \co\ and 0.77 K (Orion A) and  0.85 K (Orion B)
for the \coa\ data. 

\begin{figure*}
\begin{center}
\epsfxsize=18cm\epsfbox{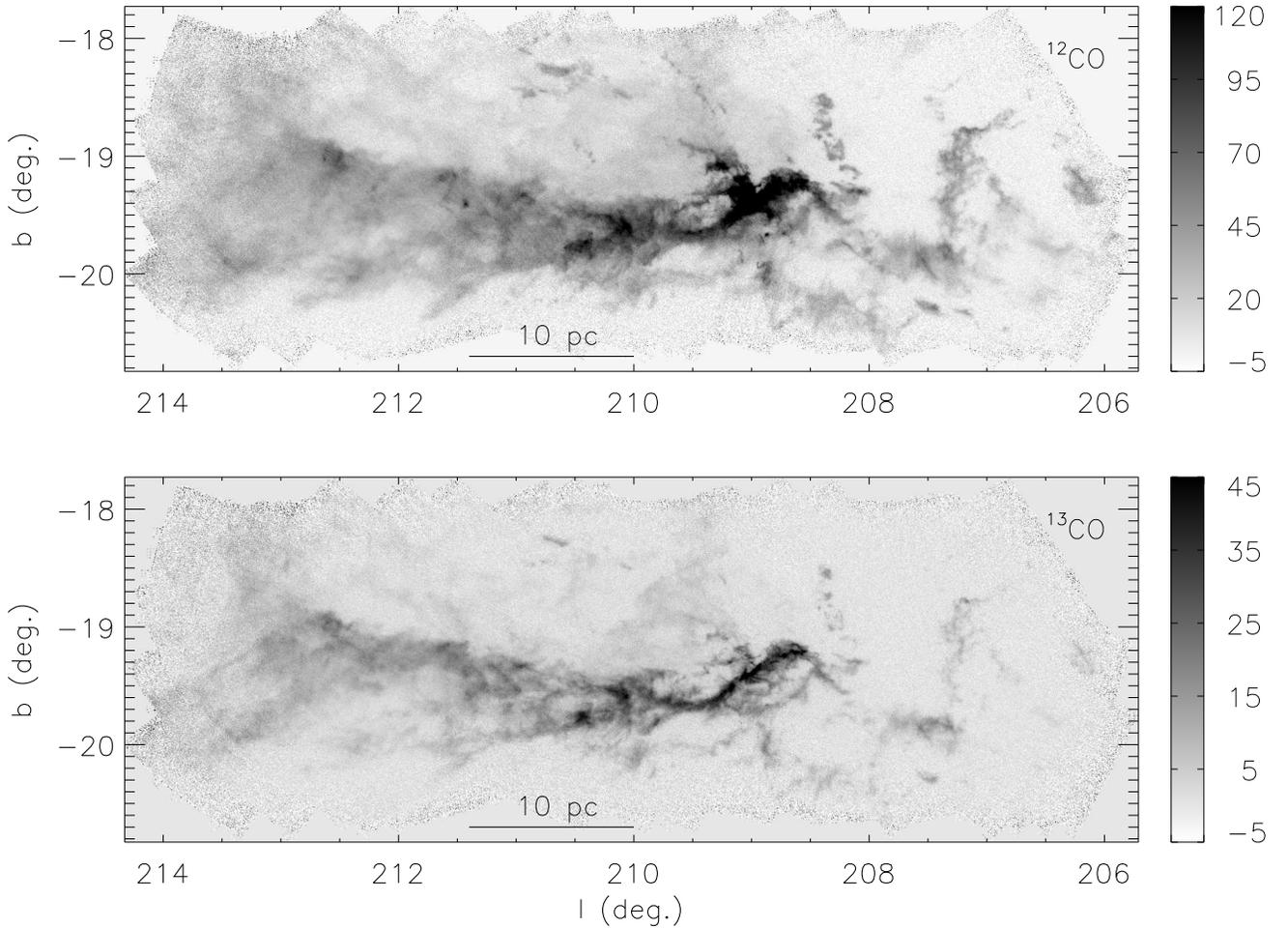}
\caption{Images of (top)  \co\  and (bottom) \coa\ J=1-0 emission from the Orion A molecular cloud integrated over \vlsr\ 0 to 16 km/s at the full resolution of the observations. Units of the colour wedges are $Kkm s^{-1}$.}
\label{fig1}
\end{center}
\end{figure*}

\begin{figure*}
\begin{center}
\epsfxsize=18cm\epsfbox{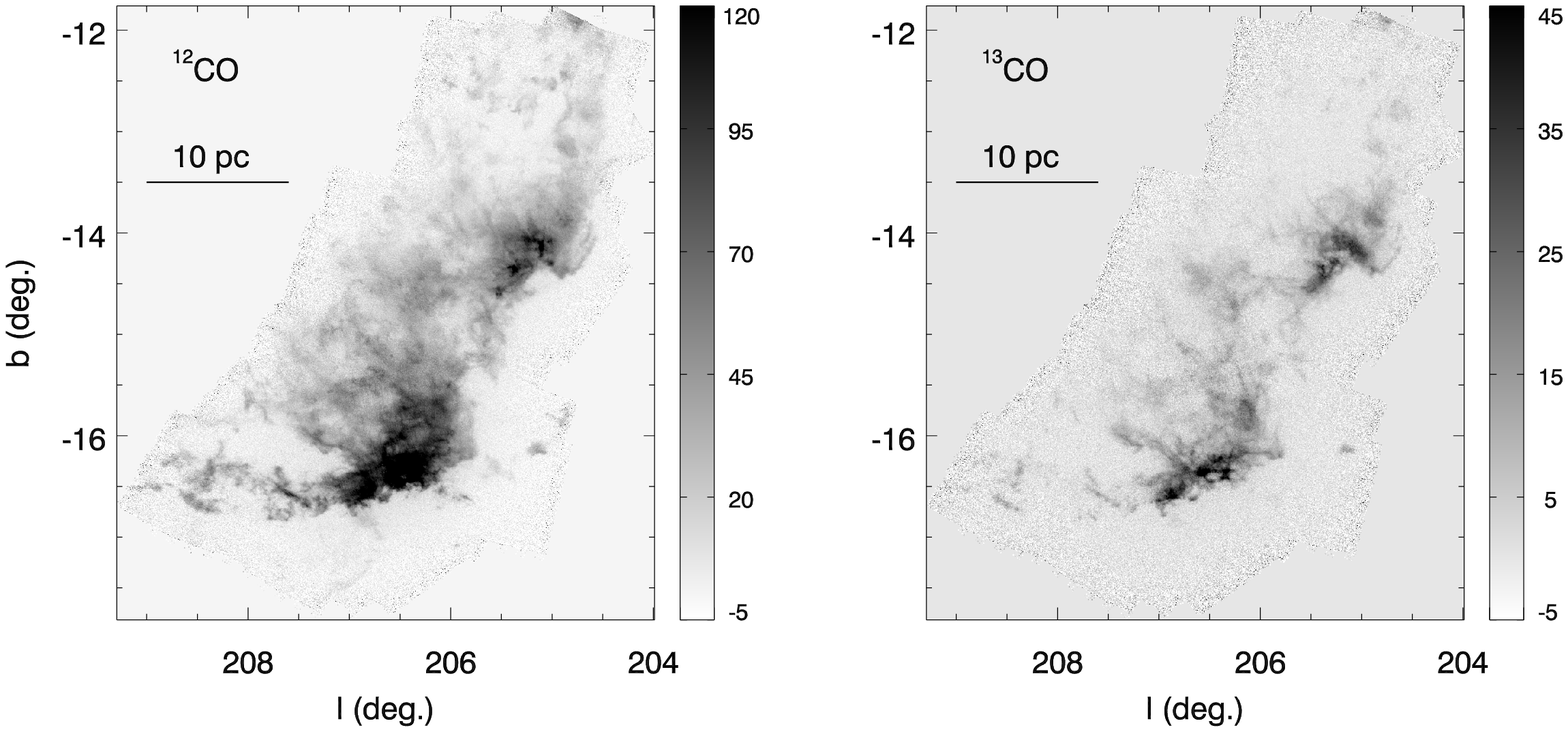}
\caption{Images of  (left) \co\ and (right) \coa\ J=1-0 emission from the Orion B molecular cloud integrated over \vlsr\ 0 to 16 km/s at the full resolution of the observations.  Units of the colour wedges are $Kkm s^{-1}$.}
\label{fig2}
\end{center}
\end{figure*}

Images of velocity integrated \co\ and \coa\ J=1-0 emission from the Orion~A and Orion~B clouds at the full angular resolution  are shown in Figure~\ref{fig1}
and Figure~\ref{fig2} respectively.  The irregularly shaped boundaries result from the scanning method.  Data near the 
boundaries exhibit higher noise owing to less integration time. 
The images show several common emission features typical of giant molecular clouds.
Regions of bright, high contrast emission are typically sites of the most recent star formation.  Such regions are often warm ($T_k > 25$ K), filamentary, and characterized by high column densities ( \nhtwo\ $>$ 10$^{22}$ cm$^{-2}$). In addition, there is an extended, low surface brightness component that comprises most of the projected area of 
the cloud.  
Recent wide field imagery of \co\ and \coa\ 
J=3-2 emission from 
Orion~A shows similar, large scale features \citep{buckle2009,buckle2012}. 
Our wide field data on Orion~B offers a significant improvement in sensitivity, 
angular and spectral resolution over 
previous maps of CO emission \citep{bally1991} 

\subsection{Near-infrared Photometry}

Near-infrared extinctions are determined from colours of stars listed in the Point Source Catalog of the Two Micron All-Sky Survey (2MASS) \citep{skrutskie2006}.  The observations are comprised of three near-infrared bands: J (1.24$\mu$m), H (1.66$\mu$m), and $K_s$ (2.16$\mu$m). The 2MASS Point Source Catalog achieves 99\% differential completeness limits of 15.8, 15.1, and 14.3 magnitudes for J, H, and K$_S$ bands respectively with 5\% photometric accuracy for bright sources and 0$\farcs$5 positional accuracy relative to the International Celestial Reference System (ICRS).  Positions and photometric information are extracted from the 
2MASS Point Source Catalog  for objects that coincide  with the area covered by 
the molecular line observations shown in  Figure~\ref{fig1} and Figure~\ref{fig2}.  The Orion A and 
Orion B fields contain 
164,930 and 242,931 sources respectively.   
 In addition, six circularly shaped control  fields with radius of 1$^{\circ}$ are analysed.  These fields are selected based 
on their large angular displacement from known regions of molecular emission.  The ($l,b$) centre positions of each control field are: (204, -22), (220,-20), (212,-23), (214,-16), (215,-23), and (220,-16). Only stars with detections in all three bands and with photometric errors less than 0.03 magnitudes are  used for the control field, resulting in a total of 26,927 stars.

%\subsection{H{\sc i} 21~cm Line Emission}
%To independently trace the atomic gas component within the Orion region, we examined the  H{\sc i} 21~cm line emission observed within the Leiden/Dwingeloo Survey 
%(Hartmann \& Burton 1997).  The angular resolution of the survey is 36\arcmin.  
 
\section{Analysis}

\subsection{Gas Excitation and \coa\ Column Density}
Methods to derive column densities of \coa\ from observations of the low rotational quantum numbers under the assumption of Local Thermodynamic 
Equilibrium (LTE) are provided by \citet{dickman1978} and more recently, \citet{pineda2010}\footnote{There is a typo in equation 18 of \citet{pineda2010}.  The argument to the exponential function should read $J (J+1)/kT_{x}$.}.  
In brief, the excitation temperature that characterizes the 
population levels of the lower and upper rotational states of \coa\ is determined 
from the peak brightness temperature of the optically thick \co\ line, $T_{12,pk}$.  For the J=1-0 transition, 
\begin{equation}
T_{ex}= {{5.53}\over{ ln(1 + {{5.53}\over{T_{12,pk}+0.83}})}}
\end{equation}
The opacity of the \coa\ emission, $\tau_{13,1}(v)$,  in the upper, J=1  rotational state is derived from the \coa\ brightness temperature, $T_{13}(v)$, 
\begin{equation}
\tau_{13,1}(v)=-ln\bigg[ 1-{{T_{13}(v)}\over{5.29}} ( [e^{5.29/T_{ex}}-1]^{-1} -0.16)^{-1}\bigg]
\end{equation}
The column density of \coa\ within the J=1  state, $N_1$, is given by the expression
\begin{equation}
N_1 = {{ {8\pi k \nu^2} \over {hc^3A_{10} }} \bigg[{ {\int \tau_{13,1}(v) dv} \over {\int (1-exp(-\tau_{13,1}(v))dv }}\bigg]
\int T_{13}(v)dv }
\end{equation}
where $A_{10}$ is the Einstein A coefficient and $\nu$ is the frequency of the J=1-0 transition.   To account for column density in the other rotational states, the assumption of 
LTE is applied,
\begin{equation}
N(^{13}CO) = N_1 {{Z}\over{2J+1}} e^{-hB_\circ {{J(J+1)}\over{kT_{ex}}}}
\end{equation}
where $B_\circ$ is the rotational constant for \coa, and Z is the partition function,
\begin{equation}
Z = \sum_{J=0}^\infty (2J+1) e^{-hB_0 {{J(J+1)}\over{kT_{ex}}}}
\end{equation}

The uncertainties in the \coa\ column densities are dominated by the assumptions of LTE. 
For regions with densities less than the critical density, the \coa\ 
rotational energy levels may be subthermally excited.  Thus, 
the 
excitation temperature derived from the more thermalized \co\ J=1-0 line 
may overestimate the excitation temperature of \coa.  
To illustrate this effect, we solve for the population in the \coa\ rotational energy levels 
using an escape probability radiative transfer method with the following conditions:
volume density of 750 \cc, \htwo\ column density of 3$\times$10$^{21}$ \cmsq, a kinetic 
temperature of 10 K, a \coa\ to \htwo\ abundance of 2$\times$10$^{-6}$,
and a velocity dispersion of 0.4 \kms. 
The model 
excitation temperatures for the 
J=1-0 transition are 9.8 K for \co\ and  5.5 K for \coa. 
In this case of subthermal excitation, 
the optical depth correction for \coa\ is underestimated leading to a
smaller column density in the J=1 level.   
This error is partially compensated by the over-correction 
for population of 
the higher excited states using a higher excitation temperature. 
For a spectrum with a peak antenna temperature of 1.5 K, 
the \coa\ LTE column density derived using the \co\ excitation 
temperature of 9.8 K is 0.7 times the column density using $T_{ex}$=5 K. 

\begin{figure*}
\begin{center}
\epsfxsize=18cm\epsfbox{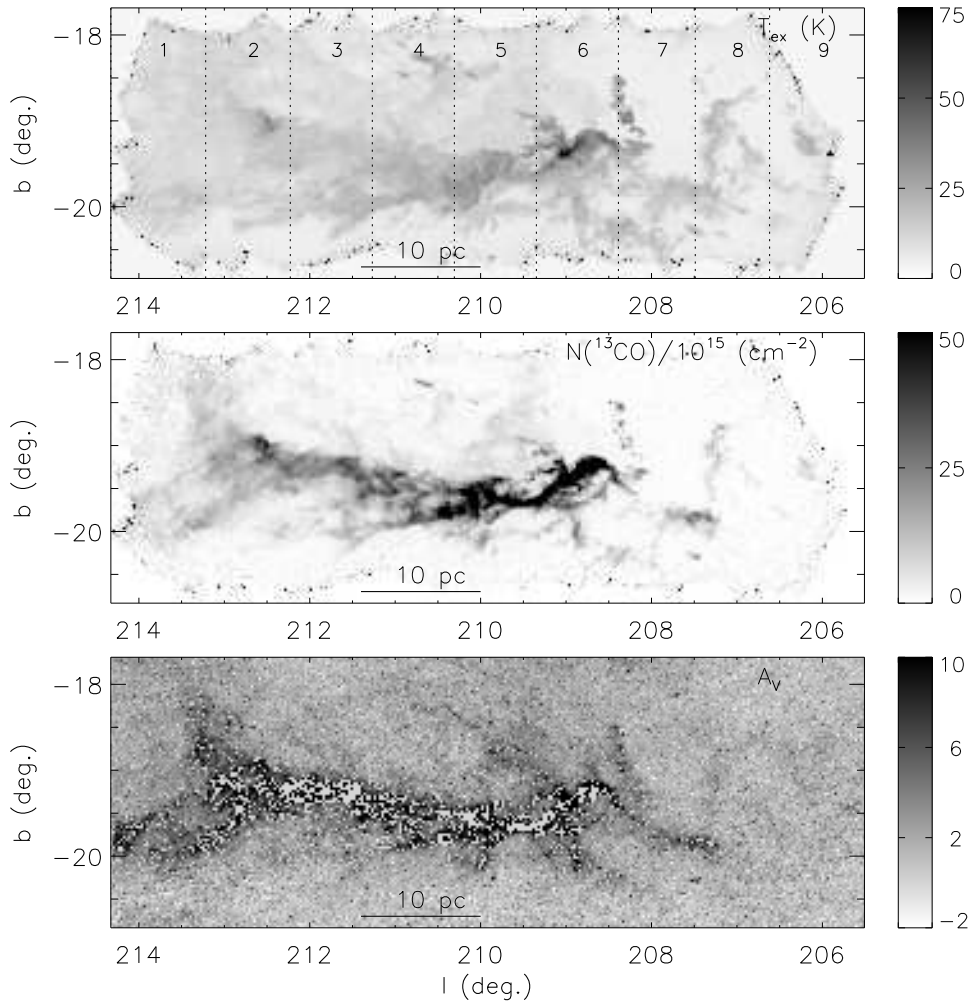}
\caption{Images of (top) gas excitation temperature, (middle) \coa\ column density,  and (bottom) infrared derived visual extinction after applying the clipping algorithm for Orion A. Pixels for which the extinction is not defined are blanked and shown as white coloured pixels. 
These are located in the high column density regions of the cloud.  The vertical dotted lines and numbers  in the $T_{ex}$ image denote the partitions discussed in the text.}
\label{fig3}
\end{center}
\end{figure*}

\begin{figure*}
\begin{center}
\epsfxsize=18cm\epsfbox{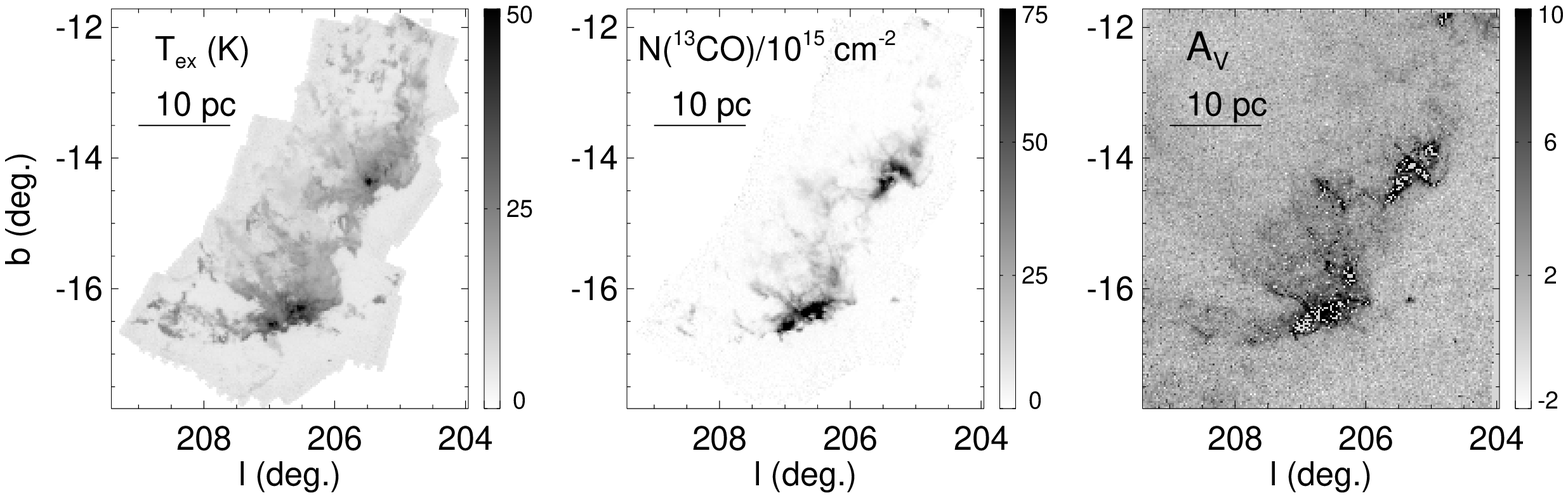}
\caption{Images of (left) gas excitation temperature, (middle) \coa\ column density, and  (right) visual extinction for Orion B. }
\label{fig4}
\end{center}
\end{figure*}

To assess the impact of the thermal noise inherent within the observations, we derive the excitation temperature and \coa\ column density using a 
Monte Carlo method that estimates the statistical errors for each quantity.  The calculations are performed 
on data cubes of the \co\ and \coa\ J=1-0 emission that are re-sampled to 1.8\arcmin\ pixel size and angular resolution. 
This resolution element corresponds to the pixel size of the infrared extinction map described in \S3.2. 
For each location on the sky, 
values of  peak \co\ brightness temperatures and \coa\ integrated intensities are extracted from Gaussian 
distributions centered on the measured value of the  peak temperature of the \co\ spectrum, $T_{12,pk}$, and $W(^{13}CO)=\int T_{13}(v)dv$ with dispersions of 
$\sigma_{12}$ and $\sigma_{13}\sqrt{ {\Delta}v{\delta}v}$ where $\sigma_{12}$ and $\sigma_{13}$ are the
root mean-square values of antenna temperatures in spectral channels with no emission,  ${\Delta}v$ is the channel spacing for each spectrum
 and ${\delta}v$ is the velocity width over which the \coa\ spectrum is integrated.  In this study, we consider 
the \vlsr\ interval of 0 to 16 \kms\ so  ${\delta}v$=16 \kms.   A given set of $T_{12,pk}$ and $W(^{13}CO)$ values are propagated through equations 1-5 to produce realizations of $T_{ex}$ and \ncoa.   This 
step is repeated 1024 times to generate distributions of $T_{ex}$ and \ncoa\ values for each pixel.
The value of the excitation temperature and statistical error at a given position, {\it x,y}, is respectively, the mean and dispersion of the $T_{ex}$ distribution.  Similarly, the mean and dispersion of the \ncoa\  distribution are assigned to the \coa\  column density and its statistical error respectively. Images of the excitation temperature and \coa\ column density are shown in Figure~\ref{fig3} for Orion A  and Figure~\ref{fig4} for Orion B.  Median 1-sigma uncertainties for $T_{ex}$ and \ncoa\ in Orion A are 0.59 K and 5$\times$10$^{14}$ cm$^{-2}$ respectively. 
For Orion B, these values are 0.55 K and 6$\times$10$^{14}$ cm$^{-2}$. 

%In practice, the \coa\ excitation temperature may not be equivalent to the \co\ excitation temperature.  Owing to the differential opacities, the \co\ excitation could be higher than that of the \coa\  in the diffuse regions of a cloud, where radiative trapping can maintain a high  level \co\ excitation.   If the true excitation temperature of \coa\ is overestimated by using the \co-derived value, then the corresponding LTE derived \coa\ column density overestimates the true column density value. 

\subsection{Visual Extinction}
The 2MASS Point Source Catalog is a powerful set of data to construct extinction images of nearby interstellar clouds. As an all sky survey, it covers the wide area subtended by nearby clouds.  In addition, a greater number of background sources are detected in the near infrared relative to optical bands owing to the reduced opacity with increasing wavelength.  To derive extinctions to each star,  we implement the Near-Infrared Color Excess Revisited (NICER) method described by \citet{lombardi01}. This method is a multi-band, maximum-likelihood approach for determining extinction to a background star based on the near-infrared colours. This is achieved through comparison of the observed stellar colour, $c_i^{obs}$,  with the intrinsic colour, $c_i^{int}$ derived from stars within a control field assumed to have no reddening. 
With the three near infrared bands of 2MASS, two colours: $c_1 = J-H$ and $c_2 = H-K_S$, are calculated.
The inferred V-band extinction is derived from the measured colour excess as
\begin{equation}
A_{v,i}= \frac{(c_i^{obs} - c_i^{int})}{k_i}
\end{equation}
where k$_i$ represents the ratio between the colour excess and the V band extinction. 
We adopt the reddening law from \citet{rieke1985}, corresponding to  $k_1 = 1/9.35$ and $k_2 = 1/15.87$.
The NICER method has the advantage of using any number of colours depending on the data available.  Extinctions are derived in cases where photometric data are only available in two of the three bands.

The intrinsic stellar J-H and H-K$_S$ colours and colour variance of background stars are derived from stellar photometry within the control fields.  The mean colours and standard deviations for stars within the control fields are: 0.446 $\pm$ 0.158 mag and 0.112 $\pm$ 0.058 mag for (J-H) and (H-K$_S$). 
%These values for the intrinsic colour of background stars are similar to those of \citet{pineda2010}, who found 0.454 mag and 0.114 mag for (J-H) and (H-K$_S$) towards Taurus, respectively.  

For each star in the target field, the NICER procedure produces an extinction, \Av, and its variance, Var(\Av), that considers both the 
variance of colours in the control fields and the photometric errors of the target star.  The uncertainty  of these extinction estimates, $\sigma_{A_v}$ = $\sqrt{Var(A_v})$, is dominated by the variance of the intrinsic colours of control field stars rather than the photometric errors of the target star.
%%%, but the photometric uncertainty of the individual star also makes a contribution. 
The median of derived extinction uncertainty values is 1.4 magnitudes for both the Orion A and Orion B clouds. 

There are several sources of contamination to identify and remove prior to folding the derived  extinction values into an image to compare with the \coa\ column density image.  As the Orion cloud is an active region of star formation, one expects a number of young stellar 
objects (YSOs) at various stages of pre-main sequence evolution within the observed fields.  Class I and II YSOs are associated with large infrared excesses due to the presence of a dense envelope or circumstellar disk that both attenuates the photospheric emission of the star and contributes to the near infrared flux from warm, thermally emitting dust grains.  Therefore, the extinction inferred from the infrared excess of Class I and Class II YSOs is  not representative of the column density through the ambient cloud.  In Orion A, the degree of contamination by YSOs
 is particularly acute near the Orion Nebula Cluster.  
The catalog of Class I and II YSOs generated by \citet{megeath2012} derived from longer wavelength photometry
from the Spitzer Space Telescope,
is used to remove such objects from the list of background stars.   A total of 2,456  (Orion A)  and 553  (Orion B) Class I and Class II YSOs are excluded from our analysis. 

\begin{figure*}
\epsfxsize=15cm\epsfbox{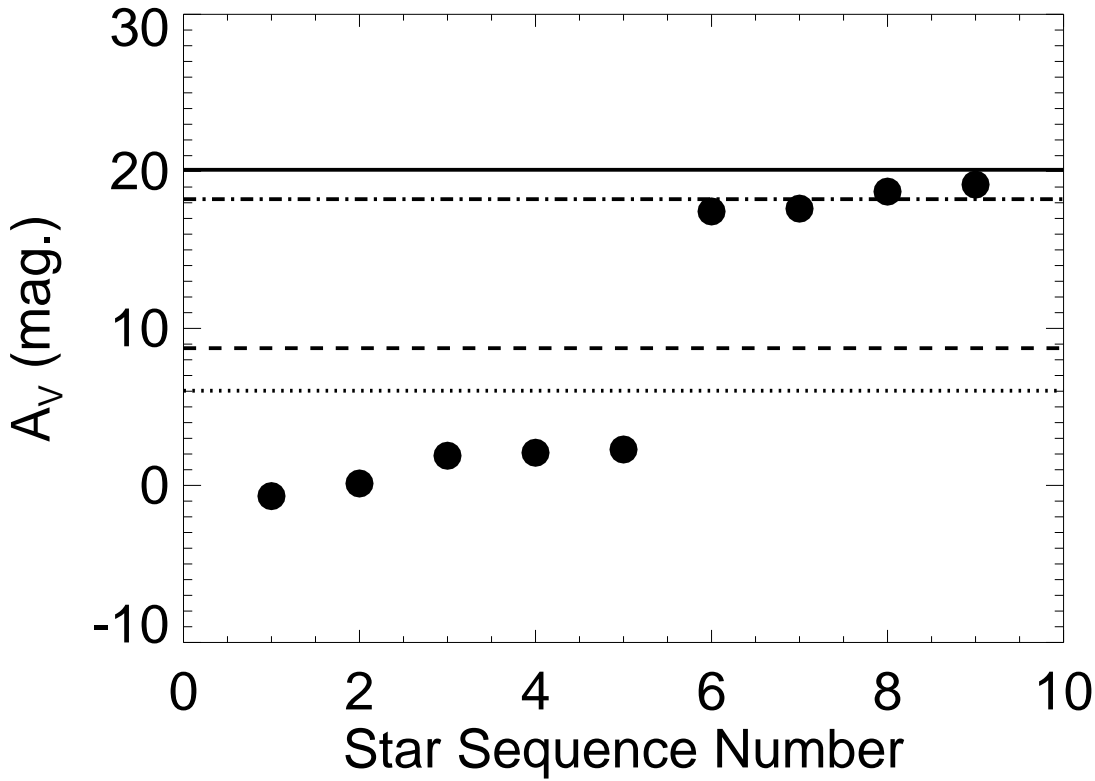}
\caption{The rank-ordered values of derived extinction for 9 stars contained within a 1.8$\times$1.8\arcmin\ area. The effect of contaminating foreground and diskless YSO sources is to lower the mean extinction. The heavy solid line represent the \Av\ value determined from \coa, the dotted line shows the threshold, $A_{v,th}$ of the clipping algorithm, the dashed lined denotes the  mean extinction value before the cut, and the dotted-dashed line shows the mean extinction value after removing the contaminating sources.}
\label{fig5}
\end{figure*}

More problematic sources of contamination are foreground stars and Class III 
YSOs within the cloud.  Both stellar types have intrinsic colours comparable to those derived in the control fields but 
whose light is not extincted by the full path of the cloud, if at all.  The effect of including such stars in deriving the local extinction is illustrated in Figure~\ref{fig5}, which shows a rare but extreme example of such contamination.  
 \Av\ values derived for  stellar sources contained within a 1.8\arcmin\ $\times$1.8\arcmin\ area of the cloud are rank-ordered for clarity.  The low extinctions determined for 5 stars is discrepant by $\sim$ 20 magnitudes with the large \Av\ values of four stars
as well as the equivalent extinction based on the \coa\ column density, 
$A_{v,eq}$=\ncoa/2.0 x 10$^{15}$  mag$^{-1}$ \citep{dickman1978}.  
%Either there is contamination by foreground and Class III YSOs or there is significant substructure in these lines of sight.  
% While 
%column density fluctuations are expected within small scales of turbulent molecular clouds due to the phenomenon of intermittency, events with amplitudes $\ge$ 20 magnitudes at these 
%size scales  (0.2 pc) should be rare (REF).  
%Yet, such large extinction differences are observed towards many locations in the  clouds.   
While such extinction differences could be due to intrinsic, 
cloud structure within the area of the pixel, 
we attribute these discrepancies to contamination by  
foreground stars and Class III pre main sequence stars that are surely present within the Orion cloud that has been 
actively forming stars for several million years.  Using extinction maps of star forming clouds at varying distances, \citet{lombardi2011} estimate 279 foreground stars per square degree towards the Orion  cloud.  This corresponds to 0.25 foreground stars per 1.8\arcmin\ sized pixel.
% or a total of 7918 stars over the area (4.8\% of the total).   
Owing to magnetic activity in their atmospheres, Class III YSOs  can be identified by their X ray emission.  
A list of X-ray identified Class III stars in Orion A (L1641 and NGC 1980) has been compiled from XMM-Newton data \citep{pillitteri2012}.  In total, 443 Class III stars from this list have 2MASS counterparts and are excluded  from our data analysis.  

The X ray survey of the Orion A cloud does not provide a complete 
census of Class III YSOs nor does it identify foreground stars.   
To isolate and remove additional Class III YSOs and  foreground stars, we apply  the 
following algorithm that conservatively leverages the additional information provided by the \coa\ column density. 
If $A_{v,eq}$ is greater than an extinction threshold, $A_{v,th} $, and the extinction of  any star within the 1.8\arcmin\  pixel is less than a 
fraction of the equivalent extinction, $f_{th}A_{v,eq}$,  then the star is excluded from any subsequent analysis. 
Optimal values for $A_{v,th} $ and $f_{th}$ are determined by maximizing the number of X-ray emitting Class III YSOs identified by the algorithm.  From this optimization, we find A$_{v,th}$ = 8 mag and $f_{th}$ = 30\%, 
which  identifies 56\% of the 443 known Class III stars.  
The algorithm  is illustrated in Figure~\ref{fig5}. 
The solid line shows the $A_{v,eq}$ estimated from \ncoa\ and the dotted line shows $A_{v,th}$ below which our algorithm would remove stars. Following the removal of the contaminating stars, the mean extinction shifts from 8.7 (dashed line) to 18.2 magnitude (dotted-dashed line).  A more liberal cut by increasing $A_{v,th}$ and/or $f_{th}$ would increase the likelihood of removal of valid background stars.  The 
algorithm is limited to regions with large \coa\ column densities.
Applying the algorithm to all stars,  2090 additional, potentially contaminating stars (1.1\% of total) are isolated and removed from the list of background stars for Orion A.  For Orion B, 824 contaminated sources are culled
from this list.  
 Contamination by foreground and/or embedded sources  may still be present in regions of lower \coa\ column density.  

An  image of dust extinction is created from the set of infrared colours of background stars within our final list following the removal of YSOs and foreground stars.   Spatial 
grids are setup that are aligned with the respective images of \coa\ column density with 1.8\arcmin\ size pixels.  The extinction of all sources whose position falls within the solid angle subtended by the pixel
are averaged with equal weighting. The resultant image has a fixed resolution throughout the map at the expense of variable noise owing to a varying number of extinction values that are averaged. 
The median 1$\sigma$ extinction errors for these sized 
pixels for \Av\ $< 10$ are 0.8 mag and 0.7 mag for Orion A and Orion B respectively.  For \Av\ $ \geq 10$,
the median extinction error is 1 mag for both clouds. These values are the propagation of the random errors for the set of extinctions that are averaged into a pixel.   The standard deviation of the set of  extinction values is typically 
larger than the associated random error.  For extinctions less than 10 mag., the median dispersions are 1.2 and 1.4 magnitudes for Orion A and Orion B respectively.  For \Av $>$ 10, the dispersions are 2.5 and 2.3 magnitudes. 
%Pixels with larger uncertainties tend to have fewer stars and higher extinctions. 
In high extinction regions, there are many pixels through which no background stars shine.  In these cases, the pixel is blanked and excluded from all subsequent analyses.  The number of blanked pixels in Orion A and Orion B are 1475 and 1002 respectively. 
 The final extinction maps constructed from the set of infrared extinctions are shown in Figure~\ref{fig3} and Figure~\ref{fig4}.  The median extinction values
over the entire map for 
Orion~A and Orion~B are  2.1 and 2.2  magnitudes respectively.  For both clouds, the extinction map 
resembles the basic outline of the \coa\ column density maps.  

\section{Results}
\subsection{\ncoa\-\Av\ Relationship}

Our analysis of near infrared stellar photometry from 2MASS and FCRAO molecular line data generates two independent views of the mass distribution in the Orion molecular cloud. 
% (see Figures~\ref{fig2} and \ref{fig5}).   
Dust extinction provides an excellent tracer of hydrogen 
column density, $N(H)=N(HI)+2N(H_2$), for a given line of sight based on the established relationships from UV spectroscopy, 
$N(H)=5.8{\times}10^{21} E(B-V)$ cm$^{-2}$-mag$^{-1}$ \citep{bohlin78,rachford2009} and $A_v/E(B-V)=3.1$ \citep{whittet2003}.   
%For fully molecular regions, $N(H_2)=9.4{\times}10^{20} A_v$ cm$^{-2}$. 
The \htwo\ column density is derived from the \coa\ column density with knowledge of the \coa\ 
abundance, \nhtwo\ = \ncoa/[$^{13}$CO/\htwo].   However, the \coa\ abundance is expected to vary with 
the local environment conditions due to changing UV radiation field intensity, gas density structure, and chemical depletion. 

\begin{figure*}
\begin{center}
\epsfxsize=12cm\epsfbox{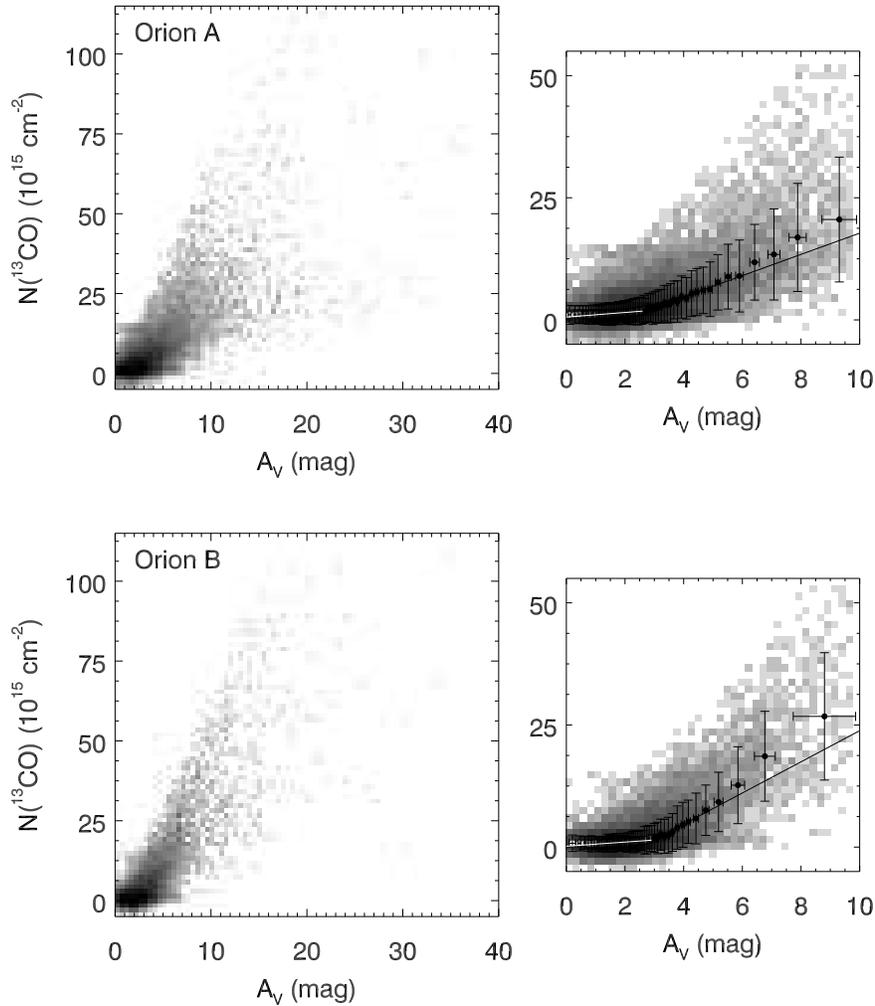}
\caption{(left) 2D histograms showing the variation of \ncoa\ with \Av\ for all defined pixels in the Orion A (top) and Orion B (bottom) clouds  over the full range 
of visual extinction.  (right)  An enlarged version of the same figure over a more limited \Av\ range.   The solid circles and vertical error bars represent the mean \ncoa\ values and standard deviation respectively within equally populated \Av\ bins. Horizontal error bars show the dispersion of \Av\ values within each bin.  The white and black solid lines show the fit of equation 7 for \Av\  $< A_{v,0}$ and 
equation 8 for 
$A_{v,0} < A_v < 10$ respectively. 
} 
\label{fig6}
\end{center}
\end{figure*}

Using the extinction as a reliable measure of the \htwo\ column density, we  assess the \coa\ abundance by examining the 
relationship between \ncoa\ and \Av. 
The variations of \coa\ column density with \Av\ for the entire Orion A and Orion B clouds are displayed in Figure~\ref{fig6} as 
shaded 2 dimensional histograms within bins of \Av\ and \ncoa.
% to determine global properties and to facilitate comparison with studies of other clouds.
%The red crosses are individual pixels in the map and the blue dots are median values of 250 pixel \Av\ bins. 
For the zoomed figures, the solid circles denote median values of \ncoa\ derived within varying sized bins of \Av\,  which contain 250 points.  The vertical and horizontal error bars reflect the standard deviation of \ncoa\ and \Av\ values respectively 
in each bin.   
Different trends are evident within three distinct regimes of extinction.  Within the first regime at low \Av\ (0 $<$ \Av\ $<$ 3 mag), \ncoa\ grows very slowly with increasing \Av. The second extinction  regime (3 $\leq$ \Av\ $\leq$ 10 mag)  is characterized by much steeper linear growth of \ncoa\ with \Av. The third regime at high extinctions (\Av\ $>$ 10 mag) shows a continued linear trend of \ncoa\ but with a much larger scatter.  For Orion A, the large scatter is due, in part, to a secondary set of points in which the \coa\ column density remains constant at the value $\sim$ 2.5$\times$10$^{16}$ \cmsq\ with increasing \Av.  
%The observed trends of the low and moderate \Av\ regimes correspond to the transition of the \coa\ abundance  from  
%the UV-irradiated envelope  and the fully, self-shielded interior of the cloud.

To quantify this transition from the low to intermediate \Av\ regime,  a non-linear least squares (Levenberg-Marquardt) algorithm is applied to simultaneously 
fit two slopes ($m_1$ and $m_2$) and an intersection point ($A_{v,int}$,$N_{int}$) to the data  over the 
range $0 < A_v \le 10$ magnitudes, 
\begin{equation}
N(^{13}CO) = m_1(\Av -A_{v,int})+N_{int}
\end{equation}
for $0 < \Av < A_{v,int}$
and 
\begin{equation}
N(^{13}CO) = m_2(\Av -A_{v,int})+N_{int} 
\end{equation}
for  $A_{v,int} \leq \Av < 10$. 
This piecewise linear equation takes into account the significantly different \ncoa-\Av\ relationship of the first two regimes. 
The fitted parameters of equations 7 and 8 for Orion A and Orion B are listed in Table~\ref{table1} and the fits are 
shown in Figure~\ref{fig6}. These values describe the observed trends in Figure~\ref{fig6} in which \ncoa/\Av\ ratio is small for \Av $< A_{v,int}$ and increases by factors of $\sim$5 (Orion A) and $\sim$8 (Orion B) for \Av\ $> A_{v,int}$. 

\begin{table*}
\centering
\caption{$N(^{13}CO)-A_v$ Relationship Fitted Parameters \label{table1}}
\begin{tabular}{cccccc}
 \hline
Partition &  $m_1$ & $m_2$ & A$_{v,int}$ & N$_{int}$ & A$_{v,3}$\\
         & (10$^{15}$ cm$^{-2}$ mag$^{-1}$) & (10$^{15}$ cm$^{-2}$ mag$^{-1}$) & (mag) & (10$^{15}$ cm$^{-2}$) & (mag.) \\
 \hline
  Orion B  & 0.42 (0.05) & 3.19 (0.20)  & 2.98 (0.05) & 1.44 (0.07) & 3.5 (0.1)\\ 
  Orion A &  0.47 (0.04) & 2.17 (0.08)  & 2.68 (0.04) & 1.84 (0.06) & 3.2 (0.1) \\ 
  1            & 0.4 (0.2)  & 1.5 (0.1) & 2.7 (0.2) & 3.0 (0.2) & 2.7 (0.2)\\ 
  2            & 0.7 (0.1)   & 2.3 (0.1) & 2.7 (0.1)  & 2.7 (0.2) & 2.8 (0.1) \\ 
  3            & 0.4 (0.2)   & 1.8 (0.1) & 2.2 (0.1) & 2.0 (0.1)  & 2.8 (0.1)\\ 
  4            & 0.5 (0.1)   & 4.4 (0.3) & 3.2 (0.1) & 2.4 (0.1) & 3.3 (0.1) \\ 
  5            & 0.8 (0.3)   & 2.0 (0.2) & 2.3 (0.2) & 2.2 (0.3) & 2.7 (0.3) \\ 
  6            & 0.5 (0.1)   & 3.6 (0.3) & 3.7 (0.1) & 2.1 (0.2) & 4.0 (0.1)\\ 
  7            & 0.3 (0.1)    & 1.7 (0.4) & 3.1 (0.2) & 1.2 (0.1) & 4.2 (0.3) \\ 
  8            & 0.1 (0.1)    & 0.9 (1.2) & 2.9 (0.8) & 0.9 (0.2) & 5.2 (3.2) \\ 
  9            & -0.07 (0.1) & 0.6 (0.5) & 2.4 (0.3) & 0.7 (0.1) & 6.2 (3.2) \\ 
 \hline
\end{tabular}
\end{table*}

\begin{figure*}
\begin{center}
\epsfxsize=15cm\epsfbox{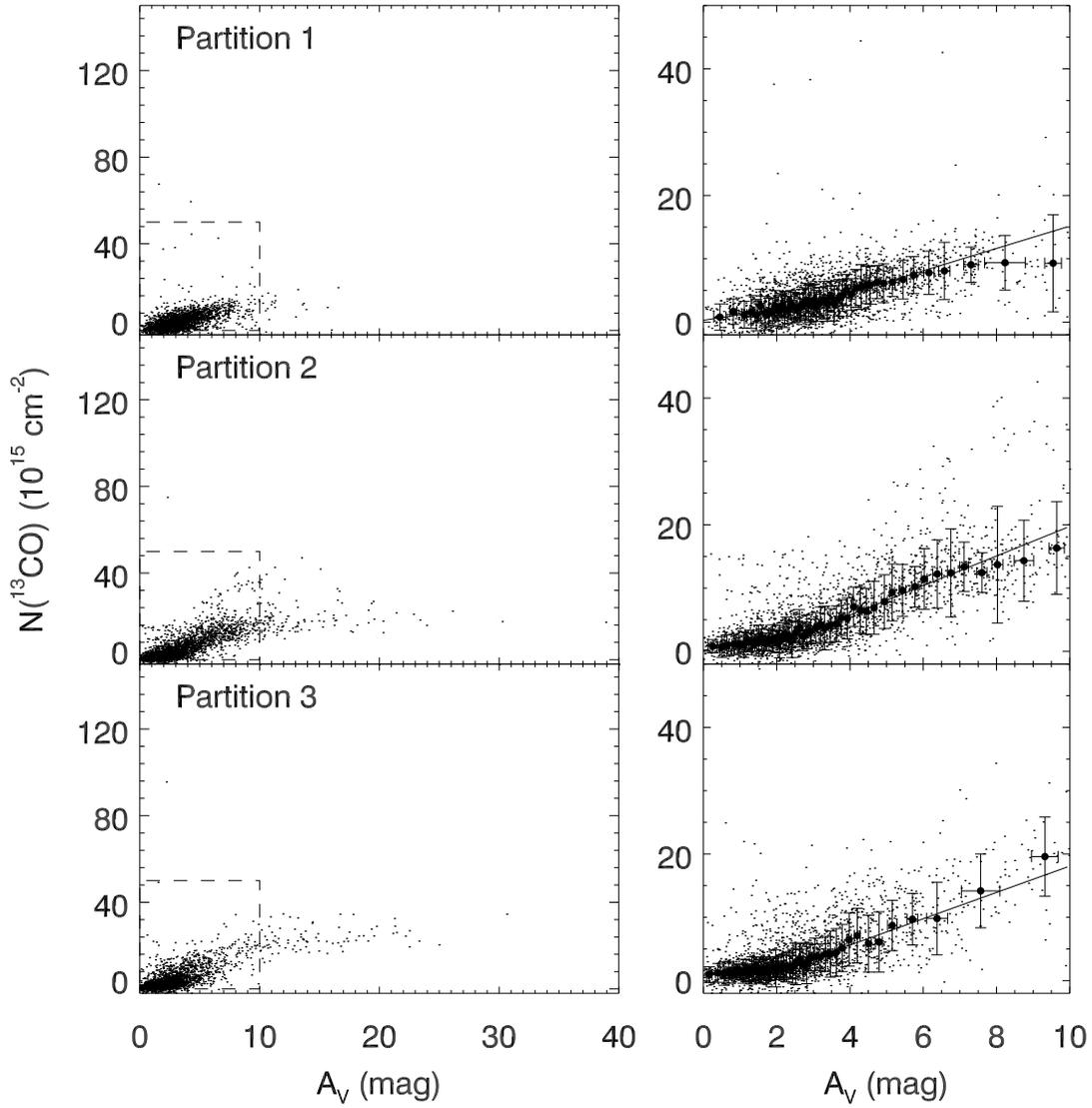}
\caption{Same as Figure~\ref{fig6} for partitions 1 (top), 2 (middle), and 3 (bottom) of the Orion~A cloud.}
\label{fig7}
\end{center}
\end{figure*}

\begin{figure*}
\begin{center}
\epsfxsize=15cm\epsfbox{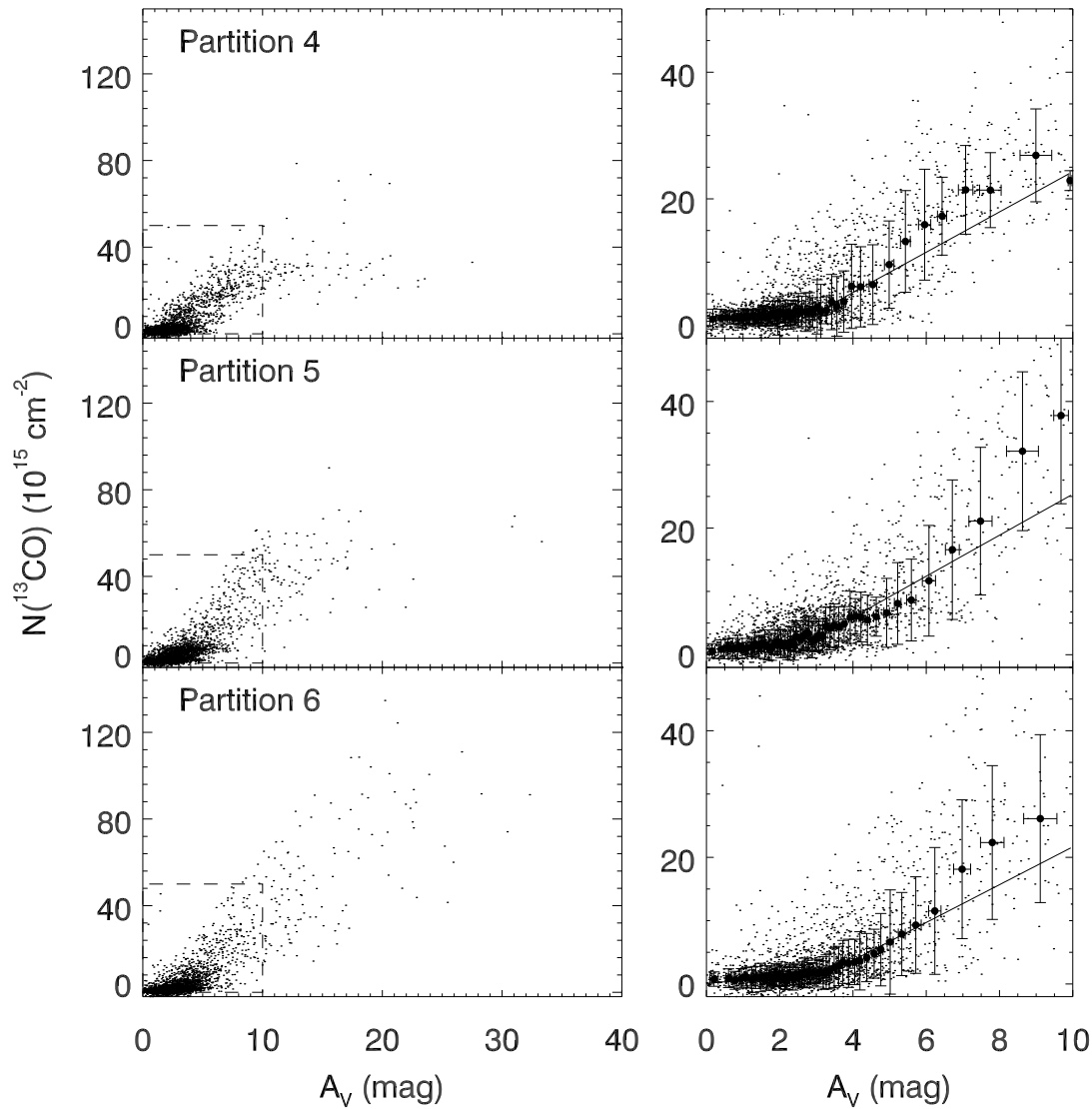}
\caption{Same as Figure~\ref{fig6} for partitions 4 (top), 5 (middle), and 6 (bottom).}
\label{fig8}
\end{center}
\end{figure*}

\begin{figure*}
\begin{center}
\epsfxsize=15cm\epsfbox{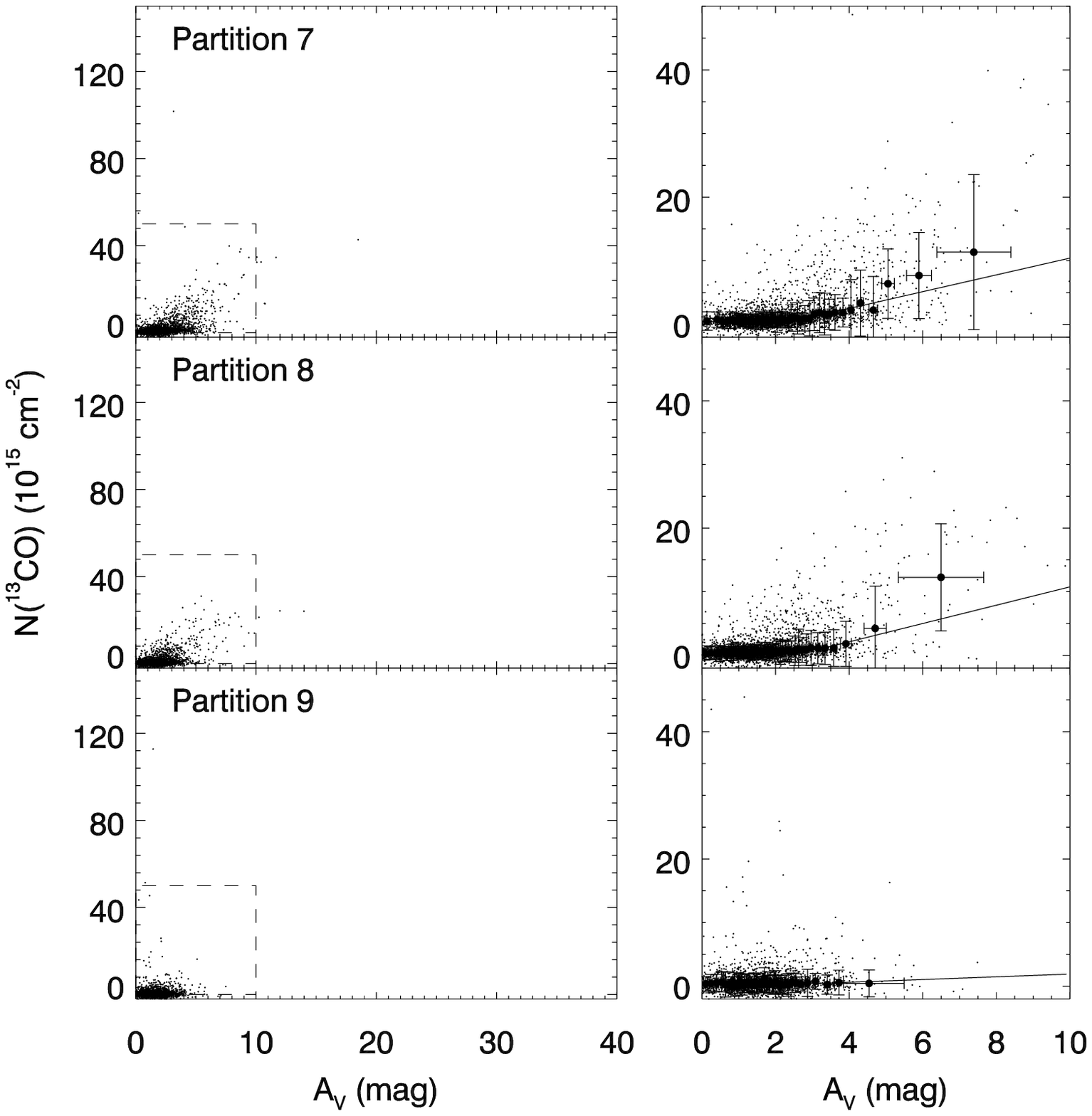}
\caption{Same as Figure~\ref{fig6} for partitions 7 (top), 8 (middle), and 9 (bottom).}
\label{fig9}
\end{center}
\end{figure*}

The Orion A cloud offers a wide range of physical conditions (FUV radiation field, gas temperature, column density) that can impact the relation between \ncoa\ and \Av.  To assess the effect of the local environment on the \ncoa-\Av\ relation, we divide the cloud into 9 Galactic longitude partitions, each containing $\sim$2,500 pixels of non-blanked data (see Figure~\ref{fig3}). The Orion Nebula, M43, and NGC 1977 are located in partition 6 and make up part of the larger Orion OB1 Association that extends from partition 5 through partition 9. 
The \ncoa-\Av\ relation for each partition is shown in Figures~\ref{fig7}-\ref{fig9}.  In these cases, the solid circles and error 
bars in the zoomed figures reflect the median and standard deviation respectively of \ncoa\ values in \Av\ bins containing 50 points. 
The trends  in the three \Av\ regimes observed in Figure~\ref{fig6} are evident in these localized regions of the cloud.  For each partition, we fit Equations 7 and 8 to the 50 point \Av\ bins for \Av $\le$ 10 magnitudes. The results from these fits are summarized  in Table~\ref{table1}.

The fitted slopes in the low \Av\ regime, $m_1$, are similar for partitions 1 through 6, ranging from 0.4-0.8$\mathrm{\times}$10$^{15}$ cm$^{-2}$ mag$^{-1}$.
The northernmost partitions (7-9) show the lowest $m_1$ values, $<$ 0.3$\rm\times$ 10$^{15}$ cm$^{-2}$ mag$^{-1}$.  These values are consistent with little or no detected \coa\ in these areas. In the intermediate \Av\ regime, the slope, $m_2$, increases by a factor of 3  from partition 1 to partition 4.   For partitions 4-6, the 2-line fit does not accurately describe the set of points in the range 5 $<$ \Av\ $<$ 10 mag in which \ncoa\ values are in excess with respect to 
the fitted line.  In these cases, the fitted parameters are leveraged by the large number of points between $A_{v,int}$ and  5 magnitudes with a lower ratio of \ncoa\ and \Av.  For regions closer to the central region of the OB1 association (partitions 7-9), the slopes rapidly decrease. 
As noted in Section 3.1, the measured \coa\ column densities may be underestimated in the low volume density and low column density 
regions of the cloud owing to 
subthermal excitation of the J=1-0 transition.  However, this error does not fully account for the 
measured differences between $m1$ and $m2$ over the expected conditions within the cloud responsible for detectable \coa\ emission. 

The third \Av\ regime also shows differential behavior.  In partitions 2-5, \ncoa\ does not continue to 
increase with increasing values of \Av\ above 
12 mag. of visual extinction.   However, in partition 6, \ncoa\ shows continued linear growth to visual extinctions of 30 mag. The different behavior in this partition accounts for the bifurcation of points in Figure~\ref{fig6} at high \Av\ values.  Partitions 1,7,8, and 9 lack sufficiently high extinction to define the behavior of \ncoa\ in this regime.

\subsection{The \wco-\Av\ Relationship}
\begin{figure*}
\begin{center}
\epsfxsize=15cm\epsfbox{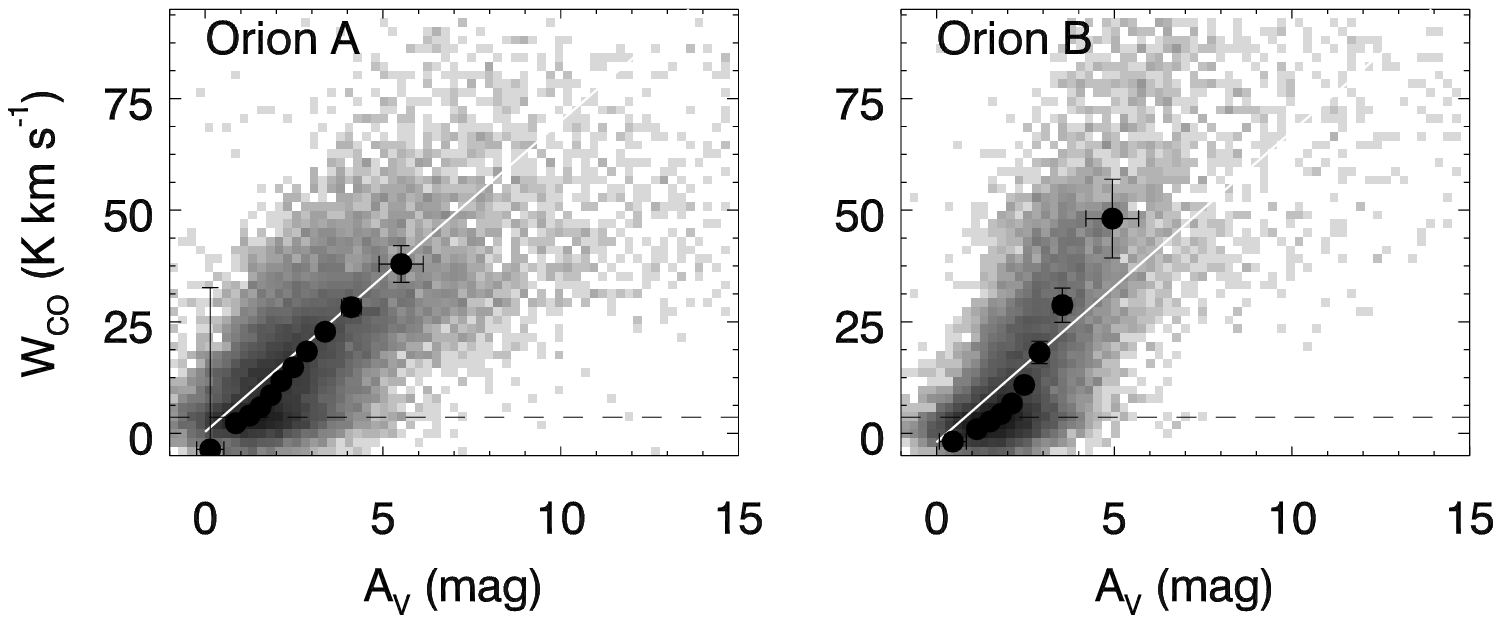}
\caption{Two dimensional histograms of \Av\ and \wco\ values for (left) Orion A and (right) Orion B.  The solid circles and error 
bars reflect the mean and standard deviation within \Av\ bins containing 2048 individual points.  The dashed horizontal line denotes 
the 3$\sigma$ threshold of the \wco. The white line shows the 
relationship implied by the standard conversion between \wco\ and \htwo\ derived from $\gamma$-rays (Strong \etal 1996). 
For both clouds, \wco\ is underluminous with respect to this standard value for \Av $<$ 2 magnitudes.  }
\label{fig10}
\end{center}
\end{figure*}

Velocity-integrated \co\ J=1-0 emission, \wco, is the most frequently applied measure of \htwo\ column density and 
mass in galaxies owing to its brightness and correlation with other tracers of \htwo\ such as $\gamma$ ray and dust emission 
in the Milky Way \citep{ade2011, bloemen1986,strong96}.  Here, we assess the point-to-point relationship of \wco\ with the infrared derived extinction in the 
Orion A and B clouds and within the Orion A partitions.  Figure~\ref{fig10} displays the variation of \wco\ with \Av\ over the 
surveyed areas of the Orion A and B clouds.  As a convenient reference, the line corresponding to the standard 
conversion of \wco\ to \htwo\ column density, $N_{H2}=1.9\times10^{20}$\wco\ cm$^{-2}/(K km s^{-1})$ and  $N_{H2}=9.4\times10^{20} A_v$ \cmsq\ is also drawn on the plots. 
As found in previous studies, \wco\ correlates well with \Av\ over the  limited 
extinction range of $\sim$2-5 magnitudes \citep{pineda2008, pineda2010}.  Yet for a fixed value of \Av, there is a large scatter of \wco\ values.  This scatter limits the use of \co\ emission as an accurate measure of the \htwo\ column density for a given line of sight 
through a molecular cloud.   For extinctions less than 2 magnitudes, there is a large number of points below the 
 reference line corresponding to weak or no detected signal of \co\ emission.   
These points correspond to gas layers in which there is significant dust extinction but 
faint, underluminous \co\ emission with respect to typical ratios of \wco\ to \Av.  
Since the derived \Av\ values are measured 
with respect to the extinction in the reference fields of background stars, the dust coupled to the atomic gas envelope 
has been largely removed.  Most, if not all, of the extinction in the range of 0.1 to 2 magnitudes, 
is linked to dust associated with molecular hydrogen. 

\begin{figure*}
\begin{center}
\epsfxsize=15cm\epsfbox{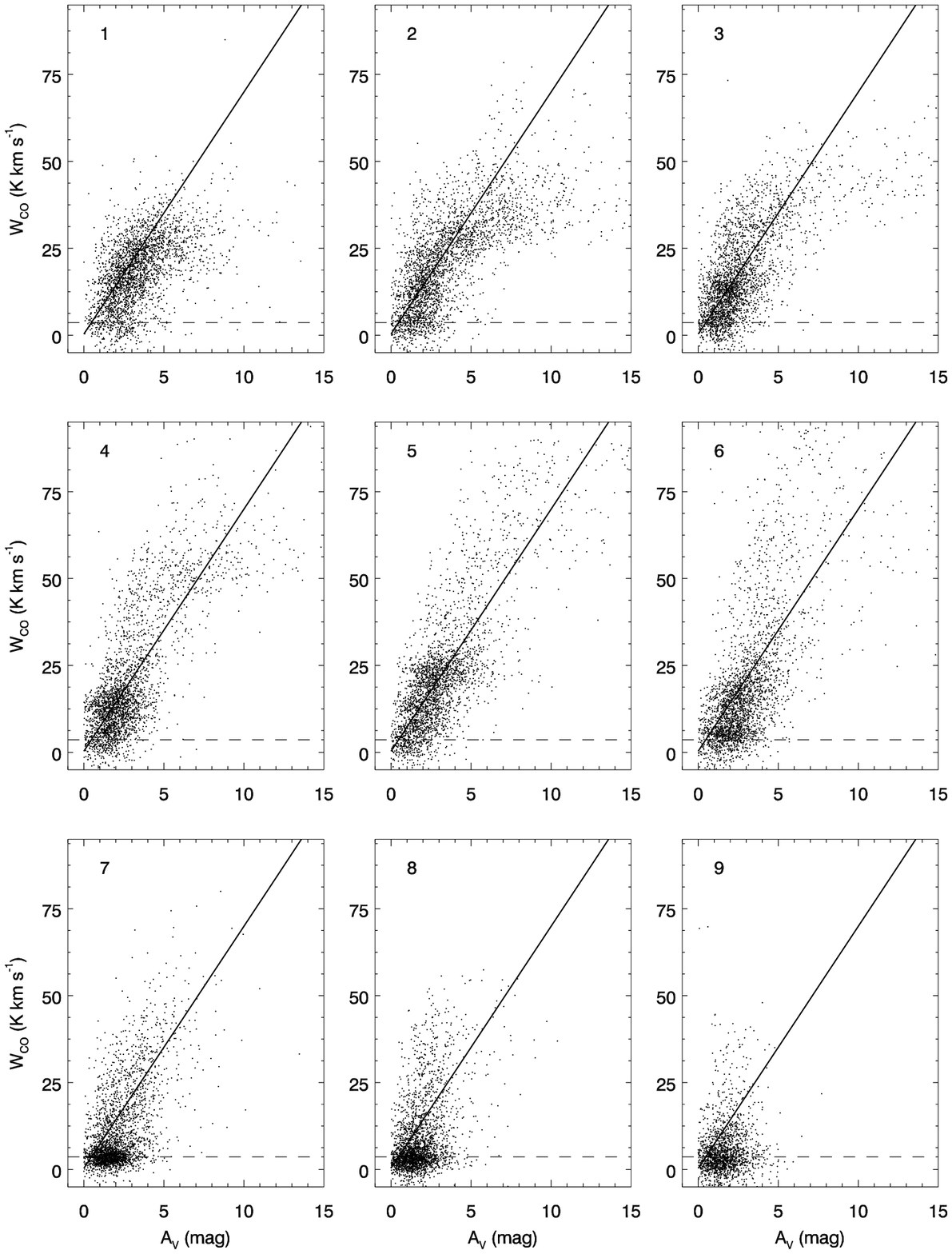}
\caption{Scatter plots of the \Av\ and \wco\ values for each Orion A partition. The  line shows the 
relationship implied by the standard conversion between \wco\ and \htwo\ and the dashed horizontal line denotes 
the 3$\sigma$ threshold of the \wco.   The overdensity of points for \Av $<$3 and \wco\ $<$ 4 K km s$^{-1}$ in partitions 6-9 reflect 
underluminous \co\ emission owing to reduced \co\ abundance in this regime. }
\label{fig11}
\end{center}
\end{figure*}

To better understand the origin of these various trends in the Orion clouds, we examine the \wco-\Av\ relationship within different environmental conditions.    Figure~\ref{fig11} shows this relationship within 
each partition of the Orion A cloud.  In the lower partitions (1-5), \wco\ linearly correlates with \Av\ between the 
range 1-5 magnitudes.  For partitions 2 and 3, \wco\ appears to saturate for \Av\ $>$ 5.  \citet{pineda2008} find a similar trend in the subregions of the Perseus molecular cloud and attribute the flattening of \wco\ with increasing \Av\  to high optical depths rather than depletion of \co\ onto dust grains.  In the areas closer to the OB association (partitions 6-9), there are many positions with \Av\ between 0 and 4 magnitudes for which CO emission is anomalously faint.  These points correspond to underluminous 
locations noted in Figure~10 but can now be isolated within the cloud in the vicinity of the Orion OB stars.

\section{Discussion}
In \S4, we identify 3 distinct regimes in the relationship between the \coa\ column density and the 
infrared-derived visual extinction.      In the following sections, we assume that the measured trends reflect the chemistry of \coa\ exposed to varying degrees of UV radiation with depth into the cloud, which in turn, depends on the local 
physical conditions of the cloud.  The abundance of \coa\ relative to \htwo\ is calculated using the derived \coa\ column densities and dust extinction as a proxy of \htwo\ column density.
These  abundances are examined in the context of photochemistry models and the changing physical conditions within the Orion  cloud.   Line excitation and optical depth of the \coa\ J=1-0 line can also impact the observed \ncoa-\Av\ relationships.  In regions of low volume density, ($n$ $<$ 10$^3$ cm$^{-3}$), that may characterize the low column density envelopes of GMCs, the J=1-0 transition is subthermally excited, ($T_{ex} <$ kinetic temperature) leading to weak emission that is difficult to detect in low sensitivity, wide field mapping programs such as the molecular line data presented in this study.   
In regions of high column density, (\Av\ $>$ 15 mag.), the \coa\ J=1-0 emission may be optically thick and our method may not fully account for this effect. 
Our limited \co\ and \coa\ J=1-0 data does not distinguish such excitation and opacity effects from chemical variations. 

\subsection{Baseline PDR Model}
To illustrate our observational results, the Meudon PDR model is used to compute the variation of \coa\ 
abundances under varying physical conditions and UV radiation fields \citep{lepetit2006}. 
The model describes the physical and chemical 
conditions within a  slab of gas and dust with hydrogen density, $n_H$, and an incident FUV radiation field strength, $\chi$, measured in units of the \citet{draine1978} radiation field.    Although the model does not consider the depletion of molecules onto grain surfaces at high densities, the photochemical calculation is sufficient for comparisons 
to data in the low to intermediate \Av\ regimes. The PDR models are calculated with parameters $\chi$ = 1, 10, 100, 10$^{3}$, and 10$^4$ at a density of 10$^3$ cm$^{-3}$ and $\chi$ = 10$^3$ and 10$^4$ at a density of 10$^4$ cm$^{-3}$. 
These values of $\chi$ and volume density bracket the expected range of physical conditions throughout most of the Orion cloud as traced by \coa\ J=1-0 emission. 
%For Orion A, one expects $\chi$ to increase with decreasing longitude corresponding to decreasing displacement from the UV 
%emitting stars of the Orion OB association.  Similarly, the volume density likely increases from the quiescent, star forming 
%environment in partitions 1-3 to the gas filaments in the vicinity of the Orion Nebula cluster where most of the current star formation 
%takes place.  
For all model parameter sets, the UV radiation field is assumed to be incident on both sides of the slab.  
It is important to note when comparing the model results 
to the observations, that the model \Av\ values are  measures of depth into the cloud with the 
corresponding \coa\ column density to each depth whereas 
the observed \Av\ and \ncoa\ values shown in Figures~\ref{fig6}-\ref{fig9}, reflect the total  extinction and column density  respectively  through the cloud.  
For each depth position in the slab as measured by \Av, the model calculates the fractional abundance of atomic and molecular hydrogen and selected molecules with respect to the total density of hydrogen protons, $n_H = n_{HI}+2n_{H2}$.  

\begin{figure*}
\begin{center}
\epsfxsize=15cm\epsfbox{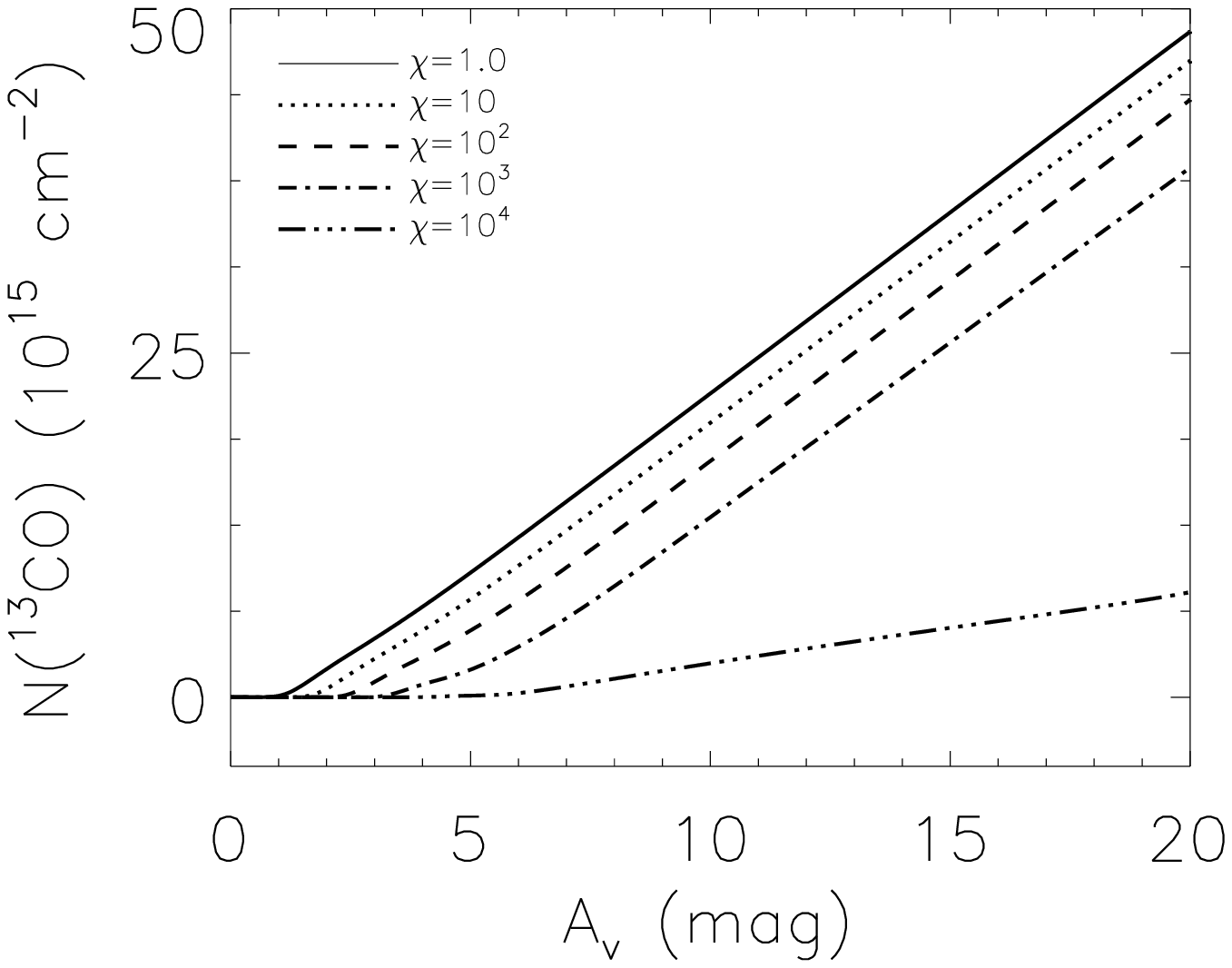}
\caption{The relation between \ncoa\ and \Av\ derived from models produced by the Meudon PDR code for  $n$=10$^3$ cm$^{-3}$ and different values of the radiation field, $\chi$ (see legend). The model does not include the effects of \coa\ 
depletion onto dust grains. }
\label{fig12}
\end{center}
\end{figure*}

The \ncoa-\Av\ relationships produced by the models with $n_H$=10$^3$ cm$^{-3}$ are shown in Figure~\ref{fig12}.  These plots exhibit similar profiles as seen in Figures~\ref{fig6}-\ref{fig9}, for the low and intermediate \Av\ ranges --- very small levels of \coa\ column density within the low \Av\ regime and a larger, yet constant value of \ncoa/\Av\ for the intermediate \Av\ regime.   For a given 
volume density, the 
depth or equivalent, model \Av\ value above which the \coa\ abundance is constant, increases with increasing UV radiation field intensity (increasing $\chi$).   A higher radiation field strength requires a larger extinction to fully self-shield \coa\ molecules to achieve a maximum abundance.    
For cloud models with n=10$^3$ \cc, the depths at which 95\% of the hydrogen is molecular are: \Av= 0.003, 0.12, 0.9, and 3.1 magnitudes for $\chi$=1,10,100,1000 respectively.  At these depths the \co\ and \coa\ abundances are several orders of magnitude smaller than their nominal, strongly self-shielded values of 1$\times$10$^{-4}$ and 2$\times$10$^{-6}$ respectively. 
%Glover \& Mac Low (2010) 
%and OTHERS/REFS note that the \htwo\ abundance is most sensitive to density and metallicity rather than the local radiation field. 
A consequence of the differential requirements for self-shielding of \htwo, \co\, and \coa\ is a layer of molecular hydrogen gas within the cloud envelope that is not traced by CO owing to its negligible abundance.  This cloud component of molecular hydrogen with no CO has been labelled ``dark'' gas by \citet{grenier2005} and may comprise a significant mass reservoir that must be considered when estimating the self-gravity  and surface density of GMCs. 

\subsection{CO Abundance Variations from UV-Illuminated Cloud Envelope to the Self-Shielded Interior}
 The low extinction regime identified in our analysis represents areas within 
the cloud where the transition to fully self-shielded \coa\ material  has not yet occurred such that photodissociation still dominates the chemistry of \coa. The shallow, non-zero slope, $m_1$, in this regime reflects the low effective abundance of \coa.   
To derive the mean relative abundance from these data, 
[\coa/\htwo] = $<$\ncoa/\Av/(\nhtwo/\Av$>$ = $m_1/ 9.4\times 10^{20}$. 
%$<N(^{13}CO)/A_v>/(N(H_2)/A_v) = m_1/ 9.4\times 10^{20}$. 
The values range from (1.1-8.5)$\times$10$^{-7}$ along the length of the Orion~A cloud.   The PDR models 
show that \coa\ abundance changes rapidly in this regime so these abundance estimates  may not be representative of  more narrow \Av\ intervals.  The models show that there is an appreciable amount of  \htwo\ mass within this layer.   

At sufficient depths into the cloud, self-shielding  enables the formation of appreciable column density of CO. In this self-shielded regime, a steeper slope, $m_2$ of the \ncoa-\Av\ relationship  is measured, relative to slopes derived in the low \Av\ regime and reflects the increased abundance of \coa. 
Within this regime, the relative abundance is [\coa/\htwo] = $ m_2/ 9.4\times 10^{20}$.
 There are significant variations of the \coa\ abundance throughout the Orion A cloud.  At the high 
longitude end of the cloud (partitions 1-3), the average 
abundance is 2.0$\times$ 10$^{-6}$.  In the central 
partitions of the cloud (partitions 4-6), the mean abundance 
increases to $\sim$3.5$\rm\times$ 10$^{-6}$.   From this peak, the 
\coa\ abundances decrease in partitions 7 and 8.  Finally, the measured slope in partition 9 is consistent with no appreciable \coa\ abundance.

These abundance results for the Orion A and Orion B molecular clouds are compared  to
values derived from other interstellar clouds with different conditions.
% with the caveat that the \ncoa\ and \Av\ data are fit differently (see equation 7) that may result in larger slopes.  
Using V band star counts to estimate \Av\ towards  38 interstellar clouds, \citet{dickman1978} found a mean relative abundance of  2.0 $\times 10^{-6}$ with a scatter of 50\%  over the range 1.5 $<$\Av$<$ 5 magnitudes.
With the availability of near infrared data to measure extinction in high column density regions, \citet{frerking82} determined the \coa\ abundances of  1.5 $\times 10^{-6}$ ( 1 $<$\Av$<$ 5) and 2.9 $\times 10^{-6}$  (4 $<$\Av$<$ 15) for the Taurus and $\rho$-Oph molecular clouds respectively.  
 For both \citet{dickman1978} and \citet{frerking82}, the abundance values are based on a limited 
number of positions in each cloud.   The advent of large-scale, multiband near infrared observations enabled 
more spatially complete estimates of molecular abundances to deeper extinction regimes.  \citet{lada94} 
investigated the dark cloud IC 5146 to \Av\ depths of 32 mag of extinction to derive a \coa\ abundance of 2.3 $\times 10^{-6}$. 
More recent studies of the \ncoa-\Av\ relationship have analysed 2MASS data.  \citet{pineda2008} derived a mean \coa\ abundance of 2.5$\times$10$^{-6}$ for the Perseus molecular cloud.  Their study also 
found $\sim$25\% variations of the \coa\ abundance with the largest abundance values in the active star forming region, NGC~1333.  \citet{pineda2010} 
found a mean abundance of 1.6$\times$10$^{-6}$ in the Taurus molecular cloud.
\citet{harjunpaa04} examined a diverse set of molecular clouds to find a broad range 
of \coa\ abundances (0.9-3.5 $\times$ 10$^{-6}$) with the largest values linked to active star forming 
regions.   

The conditions within the Orion A cloud span those found in both dark clouds as well as active star forming regions. 
The effective, global abundance  of the Orion A cloud is similar to that of $\rho$ Oph, which is a moderately 
active region of star formation containing a limited number of O and B stars.  Partitions 1-3 are typical of 
quiescent star forming regions such as Taurus and IC 5146 and exhibit  abundances similar to values found in these  dark clouds. The abundances in the warmer, star-forming regions (partitions 4-6) of Orion A are more comparable to those found in active, star-forming regions. The partitions in Orion A show a similar trend as described in \citet{harjunpaa04}. The much lower abundances found in partitions 7 and 8 are likely a result of reduced extinction and a more intense UV radiation field.   For partition 9, the second and third \Av\ regimes are non-existent, implying that the UV field intensity is too large  for \coa\ to become self-shielded. 

The elevated \coa\ abundance values in regions of active star forming regions within Orion and other targets pose an important constraint to descriptions of interstellar chemistry.   Such enhanced abundances may result from the destruction of ice mantles on 
dust grains and the release of previously depleted CO molecules.  
\citet{williams1985} summarizes several mechanisms including the evaporation of the mantles from the energy released in radical reactions  on the dust grains, photodesorption by UV radiation triggered by cosmic ray interactions with \htwo\ molecules in the interior of the cloud, and shocks generated from the interaction between dense clumps and stellar winds from young stellar objects.

\begin{figure*}
\epsfxsize=15cm\epsfbox{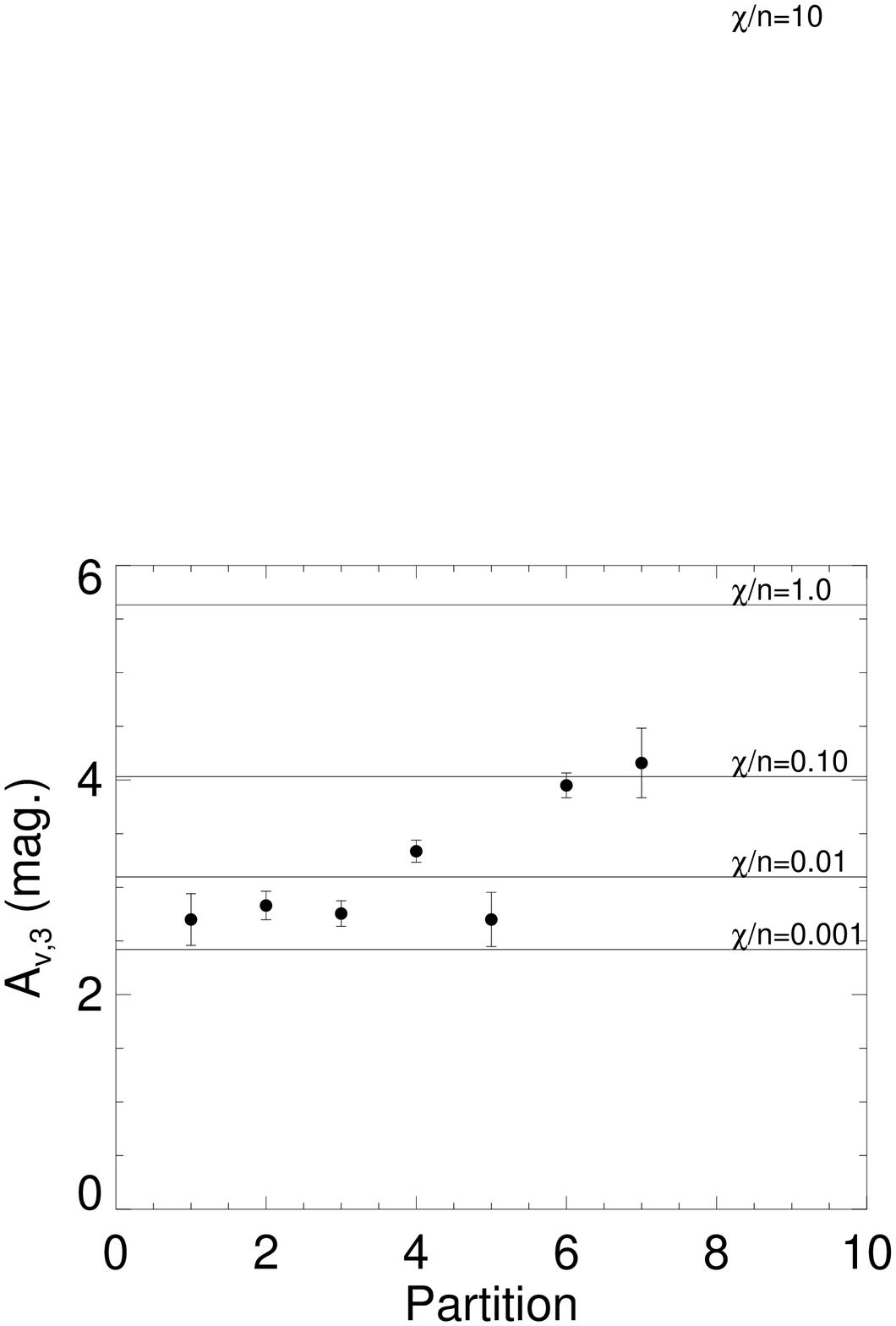}
\caption{Values of the extinction, $A_{v,3}$, at which \ncoa = 3$\times$10$^{15}$ \cmsq\ for partitions 1-7 in the Orion A cloud. The horizontal lines correspond to $A_{v,3}$ calculated for each PDR model described by 
different values of $\chi/n$.  The variation of $A_{v,3}$) in the cloud is linked to proximity to the Orion OB 
association and  changes in the local volume density of the gas. 
} 
\label{fig13}
\end{figure*}

The transition of \coa\ abundances from the UV-illuminated envelope to the self-shielded interior are dependent on the 
physical conditions.  To quantify this transition and enable a comparison of the data with models, we evaluate the \Av\ depth, $A_{v,3}$,  
at which \ncoa\ equals a nominal yet significant value of 3$\times$10$^{15}$ cm$^{-2}$.  For each partition in the Orion A cloud, 
$A_{v,3}$ is calculated  
from the parameters, $m_2$, $A_{v,int}$, and $N_{int}$.  
% For the models, $A_{v,3}$ is 
%determined by interpolating the sequence of \ncoa, \Av\ values 
%that bracket \ncoa =3$\times$10$^{15}$ cm$^{-2}$.  
Figure~\ref{fig13} shows the variation of this extinction threshold  and model-derived 
thresholds for different values of $\chi/n$.   The extinction thresholds are not well characterized in partitions 8 and 9 owing to the absence of any significant \coa\ emission and therefore, are not included in this plot. 
$A_{v,3}$ is nearly constant at $\sim$2.7 mag. in  partitions 1-3 and partition 5
 and is well bounded by models 
with 0.001 $< \chi/n <$0.005.   Given the large displacement of partitions 1-3 from the bright ionizing sources of the Orion OB association, we estimate the conditions $\chi \approx$ 1-5 and $n \approx$ 10$^{3}$ cm$^{-3}$ 
in these areas.   Partition 5 lies closer to the OB association so one might expect an enhanced UV field but this 
could be balanced by a corresponding increase in the gas density.  
For partitions 6 and 7, A$_{v,3}$ increases to 4 magnitudes and is well matched with model values with 
$\chi/n$ of $\sim$0.1.  This condition can be reasonably matched to $\chi$=10$^3$ and $n$=10$^4$ cm$^{-3}$. 
The absence of significant \coa\ emission in partitions 8 and 9, which lie near the 
centre of the OB association, likely indicates a very intense ultraviolet radiation field and a low density 
medium for which the condition for \coa\ self-shielding is not possible.  The measured changes in the threshold extinctions in the various partitions reflect the varying intensity of the 
far ultraviolet radiation field owing to displacement from the Orion OB1 association and Orion Nebula Cluster and the local volume density conditions within the cloud. 

\subsection{High \Av\ Regime}
The role of depletion of molecules onto interstellar dust grains in cold, high density regions of molecular clouds 
has been increasingly recognized owing to the availability of secondary tracers of \htwo\ column density that probe deep into the cloud, such as infrared derived extinction and sub-millimeter thermal dust continuum emission.  
 \citet{lada94} observed a flattening of \coa\ integrated intensity in the IC~5146 cloud for lines of sight with \Av\ $>$ 8-10 magnitudes.  \citet{kramer99} found a similar trend in IC 5146 for C$^{18}$O column densities for \Av\ $>$ 10, which they attributed to depletion.  Other examples of depletion include the globule B68 \citep{bergin02},  starless cores 
\citep{tafalla02}, filaments within the Taurus cloud \citep{pineda2010}. 
\citet{christie2012} examined depletion effects 
in a 
sample dust cores within several Gould's Belt clouds, including regions in Orion B.   They found depletion of \cob\ 
to be  
most severe in the Taurus cloud and smallest in the Ophiuchus cloud.   Both prestellar and protostellar cores 
within 
 the Orion~B 
cloud exhibit varying degrees of \cob\ depletion.  

Models of molecular cloud chemistry that consider depletion identify several key factors that modulate molecular abundances.  \citet{bergin95} show depletion of CO depends strongly on dust temperature, density, and timescale. 
At low densities ($<$10$^4$ \cc) and extinctions, the effect of 
depletion is small 
owing to 
desorption by cosmic rays and radiation that maintain significant fractions 
(60-90\%) of 
CO in the gas phase, even at late times of 10$^7$ years.  
For densities in excess of 10$^5$ \cc, dust temperatures of 10~K, and \Av=10 mag.,
the depletion time scales are short and much of the CO ($\sim$96\%) freezes onto the grain surfaces. 
For increasing grain temperature, molecules attached to grain surfaces  are 
released by thermal evaporation of the grain mantles.  \citet{bergin95} 
show that thermal evaporation is an extremely effective desorption process 
at dust temperatures greater than 22 K 
with the degree of 
chemical depletion depending on the effectiveness of the desorption processes. 
For temperatures larger than $\sim$25 K the models calculate that 
all CO is expected to be found in the gas phase 
However, 
temperature programmed desorption (TPD) laboratory studies demonstrate that CO molecules can be locked into water 
ice on grain mantles and that CO remains partially (25-40\%) depleted on these surfaces to even higher temperatures \citep{collings2004}. 

In the Orion A cloud, differential behavior in the high \Av\ regime is identified.  In partitions 2-5, the extinction at which the \ncoa-\Av\ relation flattens represents the depth at which environmental conditions are  appropriate for \coa\ depletion onto dust grains. The turn-over occurs at approximately 10 mag in partitions 2-4 and 15 mag in partition 5.   In contrast, partition 6 shows continued linear growth with increasing \Av\ above 10 mag.  Owing to the limited range of \Av\ in partitions 1,7,8, and 9, it is not possible to assess the role of depletion in these areas.

To examine the role of temperature in the depletion process, the gas temperatures derived from 
the optically thick CO emission in each partition are examined as a proxy or relative  measure to the grain temperature, considered 
in the models.  Table~\ref{table2} shows the mean excitation temperatures $T_{ex}$ for two regimes of extinction, moderate (5 $<$ \Av\ $< $10) and high (\Av\ $>$ 10), chosen to ensure 
that the CO lines are thermalized and to sample the high column density regions where depletion occurs.  The temperature within the high extinction regions of partitions 2-4 are all below the 22 K threshold calculated in the models, while partition 5 reaches 25 K.  In partition 6, where the temperatures are significantly higher ($T_{ex}$ = 34.0 K),   
a significant fraction of \coa\ remains in the gas phase to \Av=25 magnitudes.   
The limited amount of depletion in Partition 6 may also explain why it has the 
largest CO abundance in the cloud.   Our data can not assess whether a significant amount of CO remains 
entrapped in water ice as measured in the TPD studies.   If a large amount (50\%) 
of CO remains depleted in these regions, as measured in the TPD experiments, then 
the interstellar abundance of CO is much larger than conventionally assumed. 

Thus, CO depletion is observed in the cold, dense regions of the Orion~A cloud. 
The scatter and bifurcation of the Orion A composite \ncoa-\Av\ relation shown in Figure~\ref{fig6} is 
attributed to 
varying conditions within the cloud that allow  chemical depletion of \coa\ onto dust grains in the cold, dense regions and desorb \coa\ molecules 
in the warmer regions of the cloud. 

\begin{table*}
\caption{Properties within Partitions\label{table2}}
\begin{tabular}{cccccccccccc}
\hline
 & $T_{ex}^a$ & $T_{ex}^b$ & $M_H$ & $M_{HI}$ & $M_{H2,U}$ & $M_{H2,SS}$ & $M_{LTE,U}$ & $M_{LTE,SS}$ & L$_{CO,U}$ & L$_{CO,SS}$ & $M / L_{CO} $\\
 & (K) & (K) & (\msun) & (\msun) & (\msun) & (\msun) & (\msun) & (\msun) & (Kkms$^{-1}$pc$^2$) & (Kkms$^{-1}$pc$^2$) & (\msun/Kkms$^{-1}$pc$^2$) \\
\hline
Orion B  & 16.9 & 22.6 & 46300  &  4800 &  21900 & 19600 &    10000 & 24500 & 6100 & 8400 &  2.9 \\  
Orion A  & 19.2 & 26.3 &  63100 &  3000 &  23800 & 36200 &  15000 & 36200 & 7900 & 10800 &  3.2 \\
           1 & 12.6 & 12.9 &    8100  &   300 &    1600 &    6200 &  1400 &  4300 &  600 & 1600 & 3.6 \\
           2 & 14.3 & 15.8 &    9600 &    200 &    2000 &    7400 &   1600 & 6800 &  800 & 2000 & 3.4 \\
           3 & 15.3 & 16.9 &    7500 &    300 &    2300 &    4900 &   1900 & 4300 &  900 & 1300 & 3.2 \\
           4 & 19.6 & 21.3 &    7800 &    300 &    3000 &    4500 &   2200 & 5600 &  1100 & 1500 & 2.8 \\
           5 & 19.3 & 25.7 &    8500 &    300 &    1900 &    6300 &  1500 &  7400 &   800 & 2100 & 2.8  \\
           6 & 21.4 & 33.0 &    9300 &    600 &    3700 &    5000 &   2000 & 6900 &   1200 & 1900 & 2.8\\
           7 & 19.7 & 23.5 &    5400 &    500 &    3800 &    1200 &    1800 &  900 &   1000 & 300  & 3.6\\
           8 & 15.0 & 21.4 &    4300 &    400 &    3700 &      300 &    1600 &  300 &   1000 & 100 & 3.8 \\
           9 &   7.3 & -----  &    2600 &     100&    2500 &        13 &      900 &  10 &     400&  1  & 5.8\\
 \hline
%\multicolumn{4}{l}{% 
%\begin{minipage}{6.5cm}%
%a: For 5 $<$ \Av\ $< $10 mag \hfil \\
%b: For \Av\ $>$ 10 mag \hfil \\
%\end{minipage}%
%}\\

\end{tabular}
a: For 5 $<$ \Av\ $< $10 mag \hfil \\
b: For \Av\ $>$ 10 mag \hfil \\
\end{table*}   
\parindent=0pt

\subsection{Cloud Masses}

An important goal of this study is to assess the reliability of \co\ and \coa\ J=1-0 emission as a tracer of \htwo\ mass within 
interstellar clouds under 
varying physical conditions.  This effort is critical to studies of the molecular ISM and star formation throughout the Milky Way  and within nearby galaxies, where infrared-derived extinctions over the angular extent of 
GMCs are not possible.   Previous calibrations of the \coa\ abundance in cold, dark clouds may not apply to the 
more massive GMCs that contribute the bulk of the CO luminosity in galaxies.  The global masses of the Orion A and B clouds and the
masses within the  individual partitions of Orion A are calculated for different cloud 
layers using the derived \Av\ distributions, H{\sc i} 21~cm line emission, \co\, and \coa\ J=1-0 emission.  For the atomic gas component, we analyse the 21~cm line emission from the Atlas of Galactic Neutral Hydrogen \citep{hartmann97}.  For all mass 
calculations, we exclude positions for which the extinction is not defined in the high \Av\ regions, contain no CO data or have statistical errors for the \coa\ column density in excess of 5$\times$10$^{16}$ \cmsq.   Such 
regions account for 5.4\% and 1.6\% of the respective areas in Orion A and Orion B.   
For a given position on the sky, ${\mathbf{r}}$, 
the total hydrogen column density, $N_H({\mathbf{r}})=N_{HI}({\mathbf{r}})+2N_{H2}({\mathbf{r}})$, is derived from the infrared-derived extinction using the conversion, $N_H({\mathbf{r}})$=1.88$\times$10$^{21}$\Av(${\mathbf{r}}$) \cmsq\ \citep{rachford2009}.  The total  mass of the surveyed area is
\begin{equation}
M_{\mathrm{H}} = \mu m_{\mathrm{H}}D^2 \int_{cloud} d\Omega N_H({\mathbf{r}})
\end{equation}
where $m_{\mathrm{H}}$ is the mass of the hydogen atom, $\mu$ is the mean atomic weight to account for helium (1.36), $D$ is the distance to the cloud (410 pc for Orion), and $d\Omega$ corresponds to the solid angle of an image pixel.  
The atomic gas column density for each position is determined from the velocity integrated  H{\sc i} 21~cm line brightness temperature, $T_{21}({\mathbf{r}},v)$, 
\begin{equation}
N_{HI}({\mathbf{r}})=1.82\times10^{18}\bigg[   \int dv T_{21}({\mathbf{r}},v) \bigg] - N_{HI,ref}  \;\;\; {\mathrm{cm}}^{-2} 
\end{equation}
where $N_{HI,ref}=1.1\times10^{21}$ \cmsq\ is the atomic gas column density derived
from the  H{\sc i} 21~cm line within the reference fields used to establish the mean colours of background stars in the calculation of \Av.   
Therefore, the column densities and masses derived from \Av\ and the  H{\sc i} 21~cm line emission are relative to column density values in these reference fields.   The  H{\sc i} 21~cm line 
emission is integrated over the velocity interval -5 to 21 \kms\ to account for the more extended velocity dispersion of the atomic gas 
relative to the molecular component.  The equivalent extinction of this  H{\sc i} column density is $A_{v,HI}=N_{HI}/1.88\times10^{21}$ magnitudes. The corresponding mass of atomic gas, including helium, is 
\begin{equation}
M_{\mathrm{HI}} = \mu m_{\mathrm{H}}D^2 \int_{cloud} d\Omega  N_{HI}({\mathbf{r}}) 
\end{equation}
For the molecular gas, two components of the mass are derived from the visual extinction.  The first component is determined within 
the low \Av\ regime by considering only those lines of sight with extinctions in the range, $A_{v,HI} < A_v < A_{v,3}$, 
corresponding to where \coa\ is underabundant.    The molecular hydrogen mass within this  regime, $M_{H2,U}$,  is 
\begin{equation}
M_{H2,U} = \mu m_{\mathrm{H}}D^2 \bigg[1.88\times10^{21} \int d\Omega  Av({\mathbf{r}}) -  \int d\Omega N_{HI}({\mathbf{r}}) \bigg]
\end{equation}
for $A_{v,HI} < A_v < A_{v,3}$. 
The second mass component, $M_{H2,SS}$,  
\begin{equation}
M_{H2,SS} = \mu m_{\mathrm{H}}D^2 \bigg[ 1.88\times10^{21} \int d\Omega  Av({\mathbf{r}}) -  \int d\Omega N_{HI}({\mathbf{r}})  \bigg]
\end{equation}
for $A_v > A_{v,3}$, corresponds to  the mass within the 
strongly self-shielded layers of the cloud.   
Owing to the large difference between the beam size of the HH{\sc i} 21~cm line observations (36\arcmin) and the pixel 
size of the extinction maps (1.8\arcmin), we assume a smooth distribution of the atomic gas column density between these angular scales.  

The molecular hydrogen mass is also estimated from the distribution of \coa\ column density.  Traditionally, one assumes a constant value of the \coa\ to \htwo\ abundance ratio throughout the cloud. 
Here, we assess the errors in the \coa\ derived mass, $M_{LTE}$ using the standard \coa\ to \htwo\ abundance 2$\mathrm{\times}$10$^{-6}$ \citep{dickman1978}
that is typically applied in studies of GMCs 
\citep{carp95, heyer2009} and compare the result to the \Av\ 
derived values in both 
the underabundant and strongly self-shielded regimes. 
%%% The mass, $M_{LTE}$ for each partition is computed from the 
%image of \coa\ column density for all pixels that satisfy the condition, 
%\Av $>$A$_{v,3}$, for which a constant abundance value is appropriate.  
% A constant, standard \htwo\ to \coa\ 
%abundance value of 5$\mathrm{\times}$10$^{5}$  throughout the cloud 
%as that is typically applied in studies of GMCs, is assumed.
\begin{equation}
M_{LTE,U} = \mu m_{\mathrm{H2}}D^2 (5\times10^5) \int d\Omega N(^{13}CO)({\mathbf{r}})
\end{equation}
for $A_{v,HI} < A_v < A_{v,3}$ and 
\begin{equation}
M_{LTE,SS} = \mu m_{\mathrm{H2}}D^2 (5\times10^5) \int d\Omega N(^{13}CO)({\mathbf{r}})
\end{equation}
for $A_v > A_{v,3}$. 
Finally, we calculate the \co\ luminosity in each cloud and partition for each \Av\ regime, 
\begin{equation}
L_{CO,U} = D^2 \int d\Omega \int dv T_{12,mb}({\mathbf{r}},v)  
\end{equation}
for $A_{v,HI} < A_v < A_{v,3}$  and 
\begin{equation}
L_{CO,SS} = D^2 \int d\Omega \int dv T_{12,mb}({\mathbf{r}},v) 
\end{equation}
for $A_v >A_{v,3}$. $T_{12,mb}({\mathbf{r}},v)$ is the \co\ spectrum for position,  ${\mathbf{r}}$ within the map.  
The derived masses for the Orion A and B clouds and the 
Orion A partitions are  listed in Table~2. 
The corresponding CO luminosity to molecular mass conversion factor, $M/L_{CO}=(M_{H2,U}+M_{H2,SS})/(L_{CO,U}+L_{CO,SS})$ is also calculated. 

The atomic masses listed in Table~2 are  limited to the area covered by the CO observations and derived from atomic gas column densities from which $N_{HI,ref}$ 
has been subtracted.    
%The masses  of atomic 
%gas residing within these subtracted components are 1.7$\times$10$^4$ \msun\ and 2.0$\times$10$^4$ \msun\ 
%from the integrated areas of the Orion A and Orion B CO images respectively. 
The residual masses reflect the amount of 
atomic  material  that resides within the foreground and background layers of the Orion molecular cloud as well as the small component of atomic hydrogen expected to be present within the deep, strongly self-shielded interior 
of the cloud owing to cosmic rays \citep{goldsmith2005}.  For the lines of sight in the vicinity of the molecular cloud over the limited area shown in Figures~1 and 2, this residual atomic gas provides a small contribution (2-10\%) to the total hydrogen column density.   However, if one were to restore the layer of atomic gas represented by $N_{HI,ref}$,  the  mass of atomic gas in the observed CO area increases to 1.7$\times$10$^4$ \msun\ for Orion A and 
and 2.0$\times$10$^4$ \msun\ for Orion B.    Moreover, 
the atomic envelope of the Orion cloud  extends well beyond the surveyed area
so the total atomic mass of the Orion region is likely much larger than these calculated values.    \citet{grenier2005} tabulated the gas masses in the Orion cloud over a much larger 
area and found  a ratio of \co\ derived molecular to atomic gas mass of 1.2:1 that indicates an extended component 
of atomic gas well beyond the surveyed areas of Figures 1 and 2. 

Inspection of Table~2 shows that the \coa\ LTE column density with a constant \coa\ to \htwo\ abundance of 2$\times$10$^{-6}$  provides a reasonably accurate
measure of the \htwo\ mass residing within the self-shielded interiors of the Orion cloud despite the large abundance variations within the cloud.  
In this regime, the \coa\ mass slightly overestimates the mass determined by extinction by 25\% in Orion B.  In Orion A, the respective masses are 
identical. 
The root mean square deviation is 23\% for the 9 paritions in Orion A with a   
maximum deviation of 40\% occuring within Partition 6, 
which also exhibits an enhanced \coa\ to \htwo\ abundance ratio.     
   The self-shielded 
layer of the cloud is critical to the star formation process as it is 
the substrate from which dense filaments and cores develop  and from which newborn stars emerge.  However, this self-shielded zone comprises a fraction of the total 
molecular mass of the cloud -- 60\% in Orion A and 47\% in Orion B.   

\begin{figure*}
\epsfxsize=15cm\epsfbox{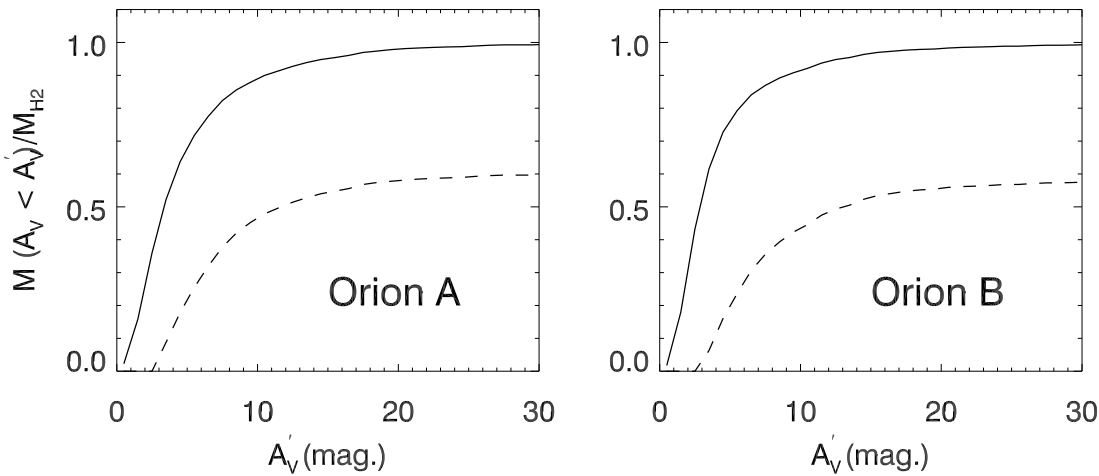}
\caption{A cumulative histogram of molecular mass in Orion A and Orion B for \Av $<$ \Av$^\prime$ as traced by \Av\ (solid line), \co, (dashed-dotted line)  assuming a CO luminosity-to-mass factor tabulated for 
each cloud in Table~2, and \ncoa\ (dashed line) with an \coa-to-\htwo\ abundance of 2.0$\times$10$^{-6}$.  50\% of the \htwo\ masses reside within the low \Av\ regimes of the clouds. 
} 
\label{fig14}
\end{figure*}

The \coa\ derived masses with a constant abundance systematically underestimate the \htwo\ mass within the low \Av, \coa\ 
underabundant regime by factors as large as 3. 
Figure~\ref{fig14} shows a cumulative histogram of the inferred \htwo\ mass, 
respectively traced by the infrared-derived \Av\ and \coa,   
which resides within varying zones of visual extinction
%In this case, $M_{H2}=M_{H2,U}+M_{H2,SS}$ as derived from \Av.  
For each cloud approximately 50\% of the \htwo\ mass resides within lines of sight with  \Av $<$ A$_{v,3}$ in which the \coa\ abundance is low and changing rapidly.  While the small opacity of the low J transitions of \coa\ are appealing to discern cloud structure and kinematics \citep{goldsmith2008,heyer2009}, investigators must recognize that these emission lines are insensitive to the 
more extended reservoir of molecular gas that may comprise a significant fraction of the cloud mass.  
 In regions of high UV radiation such as partitions 7-9 in Orion A, this missing fraction can be even larger.   

Dark gas describes the low \Av\ regime in the vicinity of molecular clouds in which molecular hydrogen is abundant due to efficient self-shielding but the abundance of \co\ is low  \citep{grenier2005, wolfire2010}.   
In these conditions, \co\ is underluminous with respect to its standard relationship with \htwo\ at moderate values of \Av.   
In the \Av\ range to which our maps are sensitive, there is little evidence for dark gas throughout 
most of the Orion A and Orion B clouds. The $M/L_{CO}$ ratios in the underluminous regime are comparable to 
those derived in the strongly, self-shielded regions.  
Dark gas is more evident in Orion A partitions 7-9 where the $M/L_{CO}$ ratios 
are larger than the values in the self-shielded regime.  
In these regions, the UV field is enhanced by the UV emitting young stars of the OB association as
illustrated in Figure~\ref{fig11} allowing for the destruction of CO molecules while maintaining a large 
fraction of molecular hydrogen gas. 
 
The conversion factor, \xco, relates the integrated \co\ emission, \wco, to the column density of \htwo\ molecules,  
$N_{H2} = X_{CO} W_{CO}$.  
The use of the optically thick \co\ line as a tracer of column density has been justified by both simple models 
of interstellar clouds \citep{dickman1986} and more recently, by computational 
simulations that include molecular chemistry \citep{glover11, narayanan12}.  

%Our comparisons of the quantities, \wco\, and \Av\ in Figures 10 and 11 show that there is indeed a correlation over a limited 
%range of extinctions, 1 $<$ \Av $<$ 5 magnitudes.  However, for a given value of \Av, the scatter of \wco\ values is large so 
%the corresponding uncertainty for the \htwo\ column density is similarly large.  
%The total \htwo\ mass is obtained 
%By integrating the column densities over the entire cloud, these  and can be related to the CO luminosity.  
%This summation averages/dilutes these large fluctutations to provide a less noisy estimate of the \htwo\ mass.  

\citet{digel1995} and \citet{digel1999} studied the \xco\ factor specifically within the Orion molecular cloud using $>$100 MeV $\gamma$-rays,
\co\, and  H{\sc i} 21~cm line emission spanning a much larger area than the more limited imaging shown in Figures 1 and 2.  
 They found \xco=1.35$\times$10$^{20}$ \cmsq/(K km s$^{-1}$) corresponding to a 
$M/L_{CO}$ ratio of 2.9 \msun/(K km s$^{-1}$ pc$^2$).  This conversion includes contributions from helium atoms to the mass budget.    
The Orion \xco\ value is lower than the Galactic mean using $\gamma$-rays \citep{strong96}
 and much lower than a recent determination using dust emission from clouds in the solar neighborhood \citep{ade2011}.  
This gamma-ray derived value of \xco\  is similar to those calculated for Orion A (3.2) and Orion B (2.9) in Table 2 using our \co\ measurements and \Av. 

More recently, the Orion cloud complex was observed by the Fermi Large Area Telescope \citep{ackermann2012}.  This study found a difference in \xco\ values 
between the high longitude end of the Orion A cloud ($l > 212$) 
where $X_{CO} = 2.3{\times}10^{20}$ cm$^{-2}$/(K km s$^{-1}$) ($M/L_{CO}$=4.9) and the remainder of the Orion A cloud as 
well as Orion B where  $X_{CO} = 1.3{\times}10^{20}$ cm$^{-2}$/(K km s$^{-1}$) ($M/L_{CO}$=2.8).  They attribute these differences to the rapid changing CO abundance in the diffuse 
gas regime at high longitudes of the Orion cloud and the larger contribution by  "dark" \htwo\ material to the total molecular mass.  While our observations do not extend as far as the 
gamma ray imagery, we derive an elevated mass weighted average $M/L_{CO}$  of  3.5 \msun/(Kkms$^{-1}$pc$^2$) within partitions 1 and 2 for which $l > 212$ 
compared to the remainder of the Orion A cloud where 
 $M/L_{CO}$=3.1 \msun/(Kkms$^{-1}$pc$^2$).  We also note the high values of $M/L_{CO}$ in partitions 7,8,9 that are close to the central region of the OB associaton and 
illuminated by local UV radiation fields.     
%Figure 11 shows weak \co\ J=1-0 emission for \Av\ $<$ 2 magnitudes in these partitions with binned values of \wco\ 
%less than \wco\ values expected when assuming the standard values of \xco\ for \Av\ $<$ 2-3 magnitudes (see Figure 10).  

Our estimates of \xco\ in Orion are smaller than the value (2.8$\times$10$^{20}$ cm$^{-2}$/(K kms$^{-1}$)) derived by \citet{paradis2012} that is also based on infrared extinctions using 2MASS.   \citet{paradis2012}
examined a more extended area than 
that shown in Figures~\ref{fig1} and \ref{fig2} and employed a different algorithm \citep{dobashi2011} to account for 
foreground star contamination, which may account for the measured differences.   

Our results for the Orion molecular cloud and those derived from gamma ray measurements emphasize that there 
are large variations of \xco\ in the interstellar medium.  The application of the 
Galactic averaged value to an individual cloud can lead to 100\% errors in the molecular mass. 
Moreover,  the scatter in the \wco-\Av\ relationship (see Figures~\ref{fig10} and \ref{fig11}) 
 demonstrate that the varying environments within a molecular cloud can 
lead to even larger errors in the \htwo\ column density.  
 
\section{Conclusions}

In this study, we derive two independent views of the molecular mass distribution in the Orion A and Orion B molecular clouds using \co\ and \coa\ line emission and IR stellar photometry from 2MASS.  These data are analysed to 
investigate the variation of \coa\ abundances over the changing physical conditions within the cloud and to evaluate the effectiveness of \coa\ emission as tracer of molecular mass in interstellar  clouds.  We find the following results. 
\begin{itemize}

\item Three distinct regimes are identified in the \ncoa-\Av\ relationship:  the photon-dominated envelope at \Av\ $<$ 3 mag., the fully self-shielded cloud interior at 3 $<$ \Av\ $<$ 10 mag. in which the \coa\ abundance is stable, and high column density zones ( \Av\ $>$ 10 mag.) in which CO can deplete onto cold dust grains. The degree to which each regime is present depends on the local physical conditions -- in particular, the strength of the UV radiation field and the temperature
of dust grains. 

\item In Orion A, the measured depth of the PDR envelope increases with proximity to the centre of the Orion OB1 association, which provides the primary source of UV radiation. This spatial dependence emphasizes the requirement of a self-shielding layer to develop significant \coa\ abundance values.  The relative \coa\ abundance within the fully, self-shielded regime varies across the cloud from a minimum of 1.4$\times$10$^{-6}$ to 3.4 $\times$10$^{-6}$.  The maximum \coa\ abundance occurs near the Orion Nebula Cluster.  

\item The presence of CO depletion onto dust grains depends on the local gas temperature that serves as a relative measure of the dust temperature. The 
observed dependence points to the role of thermal evaporation of CO molecules from grain surfaces whose temperature is above a critical value.
The regions closer to the Orion OB1 association show less depletion, which 
may account for the higher \coa\ abundances found there.

\item 
Assuming a constant \coa\ to \htwo\ abundance of 2$\times$10$^{-6}$ throughout the cloud,  
the \coa, LTE derived column densities 
provide an accurate accounting of the \htwo\ mass in layers of the 
cloud that are fully self-shielded.  In Orion, this layer comprises $\sim$50\% of the total molecular mass of the cloud.   

\item Using dust extinction as a measure of \htwo\ column density, we confirm the reduced value of \xco\ in Orion relative to the Galactic mean value previously 
derived from gamma-ray measurements.  This difference attests to variations of \xco\ in the interstellar medium. 
\end{itemize} 

\section*{Acknowledgments}
This work is supported by grants AST-0838222 and AST-1009049 from the National Science Foundation and a stipend from the Massachusetts Space Grant Consortium. 
C.~B. is funded in part by the UK Science and Technology Facilities Council grant ST/J001627/1 (“From Molecular Clouds to Exoplanets”) and the ERC grant ERC-2011- StG 20101014 (“LOCALSTAR”), both held at the University of Exeter.
This publication makes use of data products from the Two Micron All Sky Survey, which is a joint project of the University of Massachusetts and the Infrared Processing and Analysis Center/California Institute of Technology, funded by the National Aeronautics and Space Administration and the National Science Foundation.   
This research has made use of the VizieR catalogue access tool, CDS, Strasbourg, France.

\end{document}